\documentclass[aps,prd,amsfonts,amssymb,amsmath,mathrsfs,nofootinbib,reprint,showpacs,fixltx2e,subfig,usegraphicx,twocolumn]{revtex4}

\usepackage{graphicx}
\usepackage[font=small,labelfont=bf,textfont=footnotesize,justification=raggedright,singlelinecheck=true]{caption}
\usepackage{subfig}
\usepackage{aas_macros}
\usepackage{mathrsfs}
\usepackage{comment}
\usepackage{appendix}
\usepackage{multirow}
\usepackage{url}

\newcommand{\incgraph}[2]{\includegraphics[angle=270, width=#1\textwidth]{#2}}

\begin{document}

\title{Hubble without the Hubble:\\Cosmology using advanced gravitational-wave detectors alone}

\author{Stephen R. Taylor}
\email[email: ]{staylor@ast.cam.ac.uk}
\affiliation{Institute of Astronomy, Madingley Road, Cambridge, CB3 0HA, UK }

\author{Jonathan R. Gair}
\email[email: ]{jgair@ast.cam.ac.uk}
\affiliation{Institute of Astronomy, Madingley Road, Cambridge, CB3 0HA, UK}

\author{Ilya Mandel}
\email[email: ]{imandel@star.sr.bham.ac.uk}
\affiliation{NSF Astronomy and Astrophysics Postdoctoral Fellow, MIT Kavli Institute, Cambridge, MA 02139;} 
\affiliation{School of Physics and Astronomy, University of Birmingham, Edgbaston, Birmingham, B15 2TT}


\date{\today}

\begin{abstract}
We investigate a novel approach to measuring the Hubble constant using gravitational-wave (GW) signals from compact binaries by exploiting the narrowness of the distribution of masses of the underlying neutron-star population. Gravitational-wave observations with a network of detectors will permit a direct, independent measurement of the distance to the source systems. If the redshift of the source is known, these inspiraling double-neutron-star binary systems can be used as standard sirens to extract cosmological information. Unfortunately, the redshift and the system chirp mass are degenerate in GW observations. Thus, most previous work has assumed that the source redshift is obtained from electromagnetic counterparts. However, we investigate a novel method of using these systems as standard sirens with GW observations alone. In this paper, we explore what we can learn about the background cosmology and the mass distribution of neutron stars from the set of neutron-star (NS) mergers detected by such a network. We use a Bayesian formalism to analyze catalogs of NS-NS inspiral detections.  We find that it is possible to constrain the Hubble constant, $H_0$, and the parameters of the NS mass function using gravitational-wave data alone, without relying on electromagnetic counterparts. Under reasonable assumptions, we will be able to determine $H_0$ to ${\pm}10{\%}$ using ${\sim}100$ observations, provided the Gaussian half-width of the underlying double NS mass distribution is less than $0.04M_{\odot}$. The expected precision depends linearly on the intrinsic width of the NS mass function, but has only a weak dependence on $H_0$ near the default parameter values.  Finally, we consider what happens if, for some fraction of our data catalog, we have an electromagnetically measured redshift. The detection, and cataloging, of these compact-object mergers will allow precision astronomy, and provide a determination of $H_0$ which is independent of the local distance scale.
\end{abstract}

\pacs{98.80.Es, 04.30.Tv, 04.80.Nn, 95.85.Sz}

\maketitle

\section{Introduction}

The previous decade has seen several ground-based gravitational-wave (GW) interferometers built, and brought to their design sensitivity. The construction of Initial LIGO, the Laser Interferometer Gravitational-wave Observatory \citep{ligo2009,Abadie2010}, was a key step in the quest for a direct detection of gravitational waves, which are a fundamental prediction of Einstein's theory of gravity \citep{einstein1916,einstein1918}. The three LIGO detectors are located in the USA, with two sited in Hanford, Washington within a common vacuum envelope (H1, H2 of arm-lengths $4$ km and $2$ km respectively) and one in Livingston, Louisiana (L1 of arm-length $4$ km) \citep{ligo2009,Abadie2010}. The $600$ m arm-length GEO-$600$ detector \citep{geo2008} is located near Hannover, Germany.  LIGO and GEO-$600$ began science runs in 2002, and LIGO reached its initial design sensitivity in 2005. The 3 km Virgo interferometer \citep{virgo2006}, located at Cascina, Italy, began commissioning runs in $2005$, and has participated in joint searches with LIGO and GEO-$600$ since $2007$. The $300$ m arm-length TAMA-$300$ detector \citep{tama2004}, located in Tokyo, Japan had undertaken nine observation runs by $2004$ to develop technologies for the proposed underground, cryogenically-cooled, $3$ km arm-length LCGT project \citep{lcgt2010}. 


Gravitational waves from the coalescences of compact-object binaries \citep{abram1992} consisting of neutron stars (NSs) and black holes (BHs) are among the most promising sources for LIGO \citep{MandelOshaughnessy:2010}.  The first joint search for compact binary coalescence signals using the LIGO S5 science run and the Virgo VSR1 data has not resulted in direct detections, and the upper limits placed on the local NS-NS merger rate are higher than existing astrophysical upper limits \citep{Abadie2010}.  However, construction has already begun on the Advanced LIGO detectors \cite{AdvLIGO}, which are expected to increase the horizon distance for NS-NS inspirals from ${\sim}33$ to ${\sim}445$ Mpc.  This thousandfold increase in detection volume is expected to yield detections of NS-NS coalescences at a rate between once per few years and several hundred per year, with a likely value of ${\sim}40$ detections per year \cite{abadie-rate2010}.  

The Advanced Virgo detector \cite{AdvVirgo} is expected to become operational on a similar timescale as Advanced LIGO (${\sim}2015$) and with similar sensitivity. We denote the network of three AdLIGO detectors and AdVirgo as HHLV in the following. These may later be joined by additional detectors, such as LIGO Australia, IndIGO or LCGT, creating a world-wide detector network whose sensitivity will enable gravitational-wave astronomy. The network comprising LIGO-Australia, H1 and L1 will be denoted as AHL. Regardless of the precise network configuration, the hope is that when AdLIGO (and either LIGO Australia or AdVirgo) achieve their design sensitivities it will transform GW astronomy from a search for the first detection, into a tool to explore many different astrophysical and cosmological phenomena.

Gravitational waves directly encode the parameters of the emitting system, including the luminosity distance $D_L$ and the {\it redshifted} masses.  Simultaneous measurements of the redshift and the luminosity distance would allow gravitational waves to be used as standard candles, probing the cosmic distance ladder and allowing for measurements of cosmological parameters \cite{Schutz86, HolzHughes:2002}.  However, the redshift and the intrinsic masses for point-mass objects can not be individually determined from gravitational-wave observations alone.  Therefore, previous attempts to use gravitational waves as standard sirens have generally relied on the existence of electromagnetic counterparts which can be used to unambiguously measure the redshift and break the degeneracy \citep{HolzHughes:2005, nissanke2010}, or at least do so statistically \citep{macleod-hogan}.  In this paper, we demonstrate that such counterparts are not necessary if the intrinsic mass distribution is sufficiently narrow, as may be the case for double NS (DNS) binaries, although one can do even better by combining the two approaches.  

We show that it is possible to use the statistics from a catalog of gravitational-wave observations of inspiraling DNS systems to simultaneously determine the underlying cosmological parameters and the NS mass distribution.  A given cosmological model determines the redshift as a function of luminosity distance, making it possible to extract the intrinsic mass of a system from a measurement of $D_L$ and the redshifted mass. This permits us to statistically constrain the Hubble constant and the NS mass distribution via a Bayesian formalism, using only GW data. A narrower intrinsic NS mass distribution will more effectively penalize any model parameters which are offset from the true values. We investigate how the precision with which we can recover the underlying parameters scales with the number of detections and the values of the intrinsic parameters themselves. 

For the majority of our analysis, we do not consider difficult-to-detect electromagnetic (EM) counterparts to the GW detections, which were relied on in previous analyses, e.g., \citep{nissanke2010}.  Nor do we consider tidal coupling corrections to the phase evolution of DNS inspiral signals, which break the degeneracy between mass parameters and redshift to probe the distance-redshift relation \citep{messenger2011}, but which only enter at the fifth post-Newtonian order and will likely be very difficult to measure with Advanced LIGO. Rather, we rely on measurements of the redshifted chirp mass, which is expected to be the best-determined parameter, and the luminosity distance.  This approach was introduced by Markovi\'{c} in \cite{markovic1993}, where the author extracted candidate source redshifts from the redshifted chirp mass using a constant intrinsic chirp mass (this is later extended to include some spread around the assumed intrinsic value). Chernoff and Finn explored this technique in \cite{chernoff-letter-1993}, which was elaborated upon by Finn in \cite{Finn96}, where he suggested using the distribution of signal-to-noise ratios and chirp masses to probe cosmological parameters.  In this paper, we use up-to-date cosmology, mass-distribution models, expectations for detector sensitivity and parameter measurement accuracies to investigate the precision with which the Hubble constant, and NS mass distribution parameters, could be measured by the advanced GW detector network.

The paper is organized as follows. In Sec.\ \ref{sec:model}, we present a simplified analytical calculation and derive scaling laws.  Section \ref{sec:catalogue} describes the assumptions made in creating a catalog of sources, including a discussion of the DNS system properties we can deduce from a gravitational wave signal, as well as neutron-star mass distributions and merger rates.  Section \ref{sec:analysis-method} details the theoretical machinery for analyzing a catalog of detected DNS systems and the details of our implementation of this analysis.  
We describe our results in Sec.\ \ref{sec:results}, in which we illustrate the possibility of probing the Hubble constant and neutron-star mass distribution via GW data, and conclude in Sec.\ \ref{sec:conclusion} with a summary and discussion of future work.

\section{Analytical model} \label{sec:model}

Here, we present a simplified analytical model that we use to show the feasibility of our idea and to derive the main scaling relationships.  We provide additional justification for the various assumptions made in this model later in the paper.

The network of advanced detectors will be sensitive to gravitational waves from NS-NS binaries only at relatively low redshifts, $z\lesssim 0.15$ (see Sec.\ \ref{sec:properties}).  At such low redshifts, the Hubble law is nearly linear, so that to lowest order, we can write the Hubble constant as (see Section \ref{sec:properties}) 
\begin{equation}
H_0 \approx c \frac{z} {D_L}.
\end{equation}
Therefore, we expect that the uncertainty in the extrapolation of $H_0$ from redshift and distance measurements will scale as
\begin{equation} \label{eq:error}
\frac{|\delta H_0|}{H_0} \lesssim \frac{|\delta z|}{z} + \frac{|\delta D_L|}{D_L}.
\end{equation}

The detected neutron-star binaries will yield a catalog of sources with measured parameters.  These parameters will include estimates of the redshifted chirp mass ${\mathcal{M}}_{z} = (1+z)\mathcal{M}$ and luminosity distance $D_L$.  The redshifted chirp mass will be measured very accurately, so we can ignore measurement errors for this parameter.  However, our ability to extract the redshift of an individual source from the redshifted chirp mass will depend on the narrowness of the intrinsic chirp mass distribution, 
\begin{equation}
\frac{\delta z}{z} \sim \frac{\sigma_\mathcal{M}}{\mathcal{M}} \frac{1+z}{z} \sim \frac{(\sigma_\mathcal{M}/\mathcal{M})}{z},
\end{equation}
where the last approximation follows from the fact that $z \ll 1$.  On the other hand, the luminosity distance is estimated directly from the gravitational-wave signal, but with a significant error that is inversely proportional to the signal-to-noise-ratio (SNR) of the detection.

Existing binary pulsar measurements suggest that the chirp-mass distribution may be fairly narrow, $\sigma_\mathcal{M} \approx 0.06 M_\odot$ (see Sec.\ \ref{sec:mass}). Meanwhile, for the most distant sources at the threshold of detectability, $z \approx 0.15$ and $|\delta D_L| / D_L \approx 0.3$ (see Sec.\ \ref{sec:properties}).  Therefore, the first term in Eq.\ (\ref{eq:error}) is generally larger than the second term (though they become comparable for the most distant sources), and the intrinsic spread in the chirp mass dominates as the source of error. 

The errors described above were for a single detection, but, as usual, both sources of uncertainty are reduced with more detections as $1/\sqrt{N}$, where $N$ is the total number of detected binaries.  In principle, we could worry whether a few very precise measurements dominate over the remaining ${\sim}N$, affecting the overall $1/\sqrt{N}$ scaling.  The term $(\sigma_\mathcal{M}/\mathcal{M})/z$ is larger than $|\delta D_L| / D_L$, so the best measurements will be those where the former term is minimized.  The spread in the intrinsic chirp mass $\sigma_\mathcal{M}/\mathcal{M}$ is independent of the SNR. Thus, we will learn the most from measurements at high $z$, even though these will have a worse uncertainty in $D_L$ (the SNR scales inversely with $D_L$).  Therefore, somewhat counter-intuitively, the low SNR observations will be most informative.  However, since the detections are roughly distributed uniformly in the volume in which the detector is sensitive, we expect half of all detections to be within ${\sim}20\%$ of the most distant detection; therefore, we do expect a $\propto 1/\sqrt{N}$ scaling in $\delta H_0 / H_0$.\footnote{This scaling holds whenever the number of detections is increased, either because the merger rate is higher or because data are taken for longer.  On the other hand, if the number of detections increases because the detectors become more sensitive, the distance or redshift to the furthest detection will also increase, scaling with $N^{1/3}$.  In that case, as long as the first term in Eq.~(\ref{eq:error}) is still dominant, the overall improvement in $\delta H_0 / H_0$ scales as $1/N^{5/6}$.}

Using the values quoted above, for $N\sim 100$ detections, we may expect that it will be possible to extract the Hubble constant with an uncertainty of ${\sim}5\%$.  We carry out a rigorous analysis below, and find that the results of our simplistic model are accurate to within a factor of ${\sim}2$ (see Sec.\ \ref{sec:results}).

\section{Source catalog} \label{sec:catalogue}

\subsection{System properties from the gravitational waveform} \label{sec:properties}

In the following analysis we consider an advanced global network detecting the gravitational radiation from an inspiraling double-neutron-star system. The waveform for such an inspiral has a distinctive signature. Such systems are denoted \textit{chirping binaries} due to the characteristic frequency increase up to the point of coalescence.

We use the formalism of  \citep{Finn96} to describe the response of an interferometric detector to the gravitational radiation from an inspiraling binary. The detector response is a function of the system's redshifted mass, the luminosity distance to the source and the relative orientation of the source and detector. This relative orientation is described by four angles. Two of them ($\theta$, $\phi$) are spherical polar coordinates describing the angular position of the source relative to the detector. The remaining two ($\iota$, $\psi$) describe the orientation of the binary orbit with respect to the observer's line of sight \citep{Finn96}. 

In the quadrupolar approximation, the dependence of the detector response, $h(t)$, on these four angles is completely encapsulated in one variable, $\Theta$, through the equation
\begin{equation}
h(t) =
\begin{cases}
 {\frac{{\mathcal{M}}_{z}^{5/3}}{D_{L}}}{\Theta}({{\pi}f})^{2/3}{\cos}[\chi + \Phi(t)], & \text{for }t<T, \\
 0, & \text{for }t>T,
\end{cases}
\end{equation}
{\noindent}where $f$ is the GW frequency, $\chi$ is a constant phase, $\Phi(t)$ is the signal's phase, and $T$ is taken as the time of coalescence. ${\mathcal{M}}_{z}=(1+z)\mathcal{M}$ is the redshifted chirp mass, while $\mathcal{M}$ encodes an accurately measurable combination of the neutron star masses,
\begin{equation} \label{eq:chirp}
\mathcal{M} = {\left(\frac{m_{1}m_{2}}{{(m_{1}+m_{2})}^{2}}\right)}^{3/5}(m_{1}+m_{2}).
\end{equation}
Analysis of the gravitational-wave phase evolution yields errors on the deduced redshifted chirp mass which vary according to the waveform family being used. Regardless, the precision is expected to be extremely high, with a characteristic error of $\sim$0.04$\%$\footnote{${\mathcal{M}}_z$ can be determined from the strain signal in one interferometer through the phase evolution.} \citep{cutler-flanagan}. 

$\Theta$ is defined as
\begin{equation} \label{eq:angle}
{\Theta}\equiv2[F_{+}^{2}(1+\cos^{2}\iota)^{2} + 4F_{\times}^{2}\cos^{2}\iota]^{1/2},
\end{equation}
{\noindent}where $0<\Theta<4$, and
\begin{align}
F_{+}&\equiv{\frac{1}{2}}(1+\cos^{2}\theta)\cos2\phi\cos2\psi-\cos\theta\sin2\phi\sin2\psi, \nonumber\\
F_{\times}&\equiv{\frac{1}{2}}(1+\cos^{2}\theta)\cos2\phi\sin2\psi+\cos\theta\sin2\phi\cos2\psi.
\end{align}
The luminosity distance $D_{L}$ is encoded directly in the gravitational waveform; however a single interferometer cannot deduce this. The degeneracy in the detector response between $\Theta$ and $D_{L}$ must be broken and this requires a network of three or more separated interferometers to triangulate the sky location \citep{Schutz86}.
A network of separated detectors will be sensitive to the different GW polarization states, the amplitudes of which have different dependences on the binary orbital inclination angle, $\iota$. Thus, the degree of elliptical polarization measured by a network can constrain $\iota$ \citep{abram1992}. The interferometers comprising a network will be misaligned such that their varying responses to an incoming GW can constrain the polarization angle, $\psi$.

Once $\Theta$ is constrained, $D_{L}$ can then be deduced from the detector response, giving a typical measurement error of ${\sim}(300/{\rho}){\%}$, where $\rho$ is the signal-to-noise ratio of the detection (e.g., \citep{cutler-flanagan,fairhurst2011,vanderSluys:2008b}). The accuracy with which the distance can be measured will depend on the exact network configuration (for example, the introduction of an Australian detection will partially break the inclination--distance degeneracy \citep{aylott2011}), but we will use the above as a representative value.  

We don't include the impact of detector amplitude calibration errors, which could lead to systematic biases in distance estimates.  Unlike statistical measurement errors, these biases would not be ameliorated by increasing the number of detections.  For example, calibration errors of order $10\%$, as estimated for the LIGO S5 search \cite{Calibration:S5}, would translate directly into $10\%$ systematic biases in $H_0$ estimates.  Thus, systematic calibration errors could become the limiting factor on the accuracy of measuring $H_0$ if they exceed the statistical errors estimated in this paper.  


\subsection{Detector characteristics} \label{sec:detector}

For the purposes of creating a catalog of sources for our study, we are only interested in determining which binaries are detectable, and how accurately the parameters of these binaries can be estimated. We use the criterion that the network signal-to-noise ratio, $\rho_{\rm net}$, must be greater than $8$ for detection.  Actual searches use significantly more complicated detection statistics that depend on the network configuration, data quality, and search techniques, which might make our assumed detectability threshold optimistic.  Here, we are interested only in a sensible approximation of the detectability criterion.  

The network configuration for the advanced detector era is uncertain at present.  Possibilities include two LIGO 4-km detectors at Hanford and one at Livingston (HHL), probably sharing data with a 3-km European Virgo detector (HHLV).  Alternatively, one of the Hanford detectors may be moved to Australia (AHL or AHLV), improving the network's parameter-estimation accuracy \cite{aylott2011,Nissanke:2011}, while the Japanese detector LCGT and/or an Indian detector IndIGO may join the network at a later date. 

In the HHL configuration all of the sites are located in the United States, such that we may use the approximation of assuming the AdLIGO interferometers can be used in \textit{triple coincidence} to constitute a \textit{super-interferometer}. This assumption is motivated by the orientation of the interferometer arms being approximately parallel \citep{abbot2004}, and also has precedents in the literature \cite[e.g.,][]{oshaughnessy2010,Finn96}. However, source localization and $D_L$ determination is very poor in HHL, and would be greatly improved by the inclusion of data from Virgo or an Australian detector.

The single-interferometer approximation is less obviously valid for networks with distant, nonaligned detectors, such as AHL(V) or HHLV.  In \cite{searle2006}, the authors comment that the proposed LIGO-Australia site was chosen to be nearly antipodal to the LIGO sites such that all three interferometers in the AHL configuration would have similar antenna patterns.  Furthermore, since the same hardware configuration would be used for LIGO-Australia and AdLIGO, the noise spectra are expected to be similar \cite{PSD:AL}.  Meanwhile, Virgo does not have the same antenna pattern as the LIGO detectors, and the Advanced Virgo noise spectrum \cite{PSD:AV} will be somewhat different from the Advanced LIGO spectrum.  

In any case, precise comparisons of the sensitivity of different networks depend on assumptions about search strategies (e.g., coincident vs fully coherent searches) and source distributions (see, e.g., \cite{beau2008, searle2006, aylott2011}).  We therefore penalize our super-interferometer assumption in two different ways.  Firstly, we set the network SNR threshold to correspond to the expected SNR from three identical interferometers, as described below, rather than the four interferometers comprising the AHLV or HHLV networks.  We further penalize the HHLV network relative to the network including the more optimally located LIGO-Australia by raising the SNR threshold from $8$ to ${\sim}10$.  These increases in SNR thresholds have the effect of restricting the network's reach in luminosity distance or redshift; however, similar numbers of detections can be achieved by longer observation times.

With the aforementioned caveats, we proceed with our assumption that a global network can be approximated as a single super-interferometer. This is to provide a proof of principle for the ability of such a network to probe the background cosmology and aspects of the source distribution. We do not anchor our analysis to precise knowledge of the individual interferometer site locations and orientations, but will attempt to correct for any possible bias.

Following \cite{Finn96} (and correcting for a missing square root), we can write the matched filtering signal-to-noise ratio in a single detector as
\begin{equation} \label{eq:snr}
{\rho}=8{\Theta}{\frac{r_{0}}{D_{L}}}{\left({\frac{{\mathcal{M}}_{z}}{1.2M_{\odot}}}\right)}^{5/6}\sqrt{\zeta(f_{\rm{max}})},
\end{equation}
{\noindent}where
\begin{align}
{r_{0}^{2}}&\equiv{\frac{5}{192\pi}}{\left({\frac{3}{20}}\right)}^{5/3}x_{7/3}M_{\odot}^{2},\nonumber\\
x_{7/3}&\equiv{\int_{0}^{\infty}\frac{df({\pi}M_{\odot})^{2}}{{({\pi}fM_{\odot})^{7/3}}S_{h}(f)}},\nonumber\\
{\zeta(f_{\rm{max}})}&\equiv{\frac{1}{x_{7/3}}}{\int_{0}^{2f_{\rm{max}}}\frac{df({\pi}M_{\odot})^{2}}{{({\pi}fM_{\odot})^{7/3}}S_{h}(f)}}.
\end{align}
$S_{h}(f)$ denotes the detector's noise power spectral density (the Fourier transform of the noise auto-correlation function), and $2f_{\rm{max}}$ is the wave frequency at which the inspiral detection template ends \citep{Nutzman2004}. The SNR of a detected system will vary between the individual network sites, as a result of the different $S_{h}(f)$'s and angular dependencies. The network SNR of a detected system is given by the quadrature summation of the individual interferometer SNRs,
\begin{equation}
{\rho}_{\rm{net}}^{2}=\displaystyle\sum_{k}{\rho}_{k}^{2}.
\end{equation}
We approximate the sensitivity of the super-interferometer by assuming $3$ identical interferometers in the network with the sensitivity of Advanced LIGO, such that $r_{0,\rm{net}}\approx\sqrt{3}r_0$.  Different target noise curves for AdLIGO produce different values for $r_0$, which vary between $\sim 80-120$ Mpc \citep{PSD:AL}. We adopt the median value of $100$ Mpc for a single interferometer, yielding $r_{0,\rm{net}}\sim 176$ Mpc for the network.


The SNR also depends on ${\zeta}(f_{\rm{max}})$, which increases monotonically as a function of $f_{\rm{max}}$. This factor describes the overlap of the signal power with the detector bandwidth \citep{Finn96}, which will depend on the wave frequency at which the post-Newtonian approximation breaks down, and the inspiral ends. It is usual to assume that the inspiral phase terminates when the evolution reaches the innermost stable circular orbit (ISCO), whereupon the neutron stars merge in less than an orbital period. This gives
\begin{equation}
f_{\rm{max}}^{\rm{GW}} = 2f_{\rm{max}} = 2\left(\frac{f_{\rm{ISCO}}}{1+z}\right) = \frac{1570\text{ Hz}}{1+z}\left({\frac{2.8M_{\odot}}{M}}\right),
\end{equation}
{\noindent}where $M$ is the total mass of the binary system \citep{abadie-rate2010}. $f_{\rm{ISCO}}$ also depends directly on the mass ratio ${\mu}/M$ ($\mu$ is the system's reduced mass); however this mass asymmetry term has a negligible effect on $f_{\rm{max}}$ for the mass range of neutron stars considered here \citep{Kidder,Nutzman2004}. 
The maximum binary system mass could conceivably be ${\sim}4.2M_{\odot}$.\footnote{Both neutron stars in the binary system would need to have masses $2\sigma$ above the distribution mean at the maximum considered $\mu$ and $\sigma$, where ${\mu}_{\rm{NS}}\in[1.0,1.5]M_{\odot}$, ${\sigma}_{\rm{NS}}\in[0,0.3]M_{\odot}$.} The AdLIGO horizon distance for $1.4M_\odot$--$1.4M_\odot$ inspirals is ${\sim}445$ Mpc, which corresponds to $z\sim 0.1$ in the $\Lambda$CDM cosmology. Given that we are evaluating different cosmological parameters, we adopt $z\sim 1$ as a generous upper redshift limit to a second-generation network's reach. This redshift exceeds the reach of AdLIGO in all considered cosmologies\footnote{$H_{0}\in[0.0,200.0]$ km\ s$^{-1}$Mpc$^{-1}$; ${\Omega}_{k,0}=0$; ${\Omega}_{m,0}\in[0.0,0.5]$} and chirp masses. With these extreme choices for the variables, the orbital frequency at the ISCO, $f_{\rm{max}}$, could be as low as ${\sim}262$ Hz. For the latest {\it zero-detuning-high-power} AdLIGO noise curve \cite{PSD:AL}, ${\zeta}(f_{\rm{max}}=262 \rm{Hz})\gtrsim0.98$. Thus, we feel justified in adopting ${\zeta}(f_{\rm{max}})\simeq1$ for the ensuing analysis.

Thus matched filtering, with an SNR threshold of 8, a characteristic distance reach of ${\sim}176$ Mpc and ${\zeta}(f_{\rm{max}})\simeq1$, provides a criterion to determine the detectability of a source by our network.\footnote{There will be some bias in this approximation, since we are assuming each interferometer records the same SNR for each event. The fact that the different interferometers are not colocated means that this may overestimate the number of coincident detections. We carry out the analysis here aware of, but choosing to ignore, this bias, and in Sec.\ \ref{sec:merger-vary} consider raising the network SNR threshold, which has the same effect as reducing the \textit{characteristic distance} reach of the network.}

\subsection{Orientation function, $\Theta$} \label{sec:orient}

The angular dependence of the SNR is encapsulated within the variable $\Theta$, which varies between 0 and 4, and has a functional form given by Eq.\ (\ref{eq:angle}). From our catalog of coincident DNS inspiral detections we will use only ${\mathcal{M}}_z$ and $D_L$ for each system. The sky location and binary orientation can be deduced from the network analysis, however we will not explicitly consider them here. Without specific vales for the angles ($\theta$, $\phi$, $\iota$, $\psi$) we can still write down the probability density function for $\Theta$ \citep{Finn96}. Taking $\cos\theta$, $\phi/\pi$, $\cos\iota$ and $\psi/\pi$ to be uncorrelated and distributed uniformly over the range [$-1,1$], the cumulative probability distribution for $\Theta$ was calculated numerically in \cite{chernoff93}. The probability distribution can be accurately approximated \citep{Finn96} by,
\begin{equation} \label{eq:ptheta}
{\mathcal{P}}_{\Theta}(\Theta)=
\begin{cases}
\frac{5}{256}{\Theta}(4-\Theta)^3, & \text{if }0<\Theta<4,\\
0, & \text{otherwise. }
\end{cases}
\end{equation}
We can use Eq.\ (\ref{eq:ptheta}) to evaluate the cumulative distribution of $\Theta$, 
\begin{equation} \label{eq:ctheta}
C_{\Theta}(x)\equiv\int_{x}^{\infty}{\mathcal{P}}_{\Theta}(\Theta)d\Theta\simeq
\begin{cases}
1, & \text{if }x\leq0\\
\frac{(1+x)(4-x)^4}{256}, & \text{if }0\leq x\leq4\\
0, & \text{if }x>4.
\end{cases}
\end{equation}

\subsection{Mass distribution} \label{sec:mass}

In recent years, the number of cataloged pulsar binary systems has increased to the level that the underlying neutron-star mass distribution can be probed.  There is now a concordance across the literature that the neutron star mass distribution is multimodal, which reflects the different evolutionary paths of pulsar binary systems \cite{kiziltan2010,valentim2011}.  However, we are only concerned with neutron stars in NS-NS systems for this analysis, and their distribution appears to be quite narrow.  In particular \cite{valentim2011} found that the neutron stars in DNS systems populate a lower mass peak at $m \sim1.37M_{\odot}\pm0.042M_{\odot}$.  Meanwhile, \cite{kiziltan2010} restricted their sample of neutron stars to those with secure mass measurements and predicted that the posterior density for the DNS systems peaked at ${\mu}_{\rm{NS}}\sim1.34M_{\odot}$, ${\sigma}_{\rm{NS}}\sim0.06M_{\odot}$.

Population synthesis studies of binary evolution predict similarly narrow mass distributions for neutron stars in NS-NS binaries (see, e.g., \cite{Belczynski:2008,oshaughnessy2010,MandelOshaughnessy:2010} and references therein).  Some models predict that the mass of neutron stars at formation is bimodal, with peaks around $1.3$ and $1.8$ solar masses, and any post-formation mass transfer in DNS systems is not expected to change that distribution significantly, but the $1.8M_{\odot}$ mode is anticipated to be very rare for DNS systems, with the vast majority of merging neutron stars belonging to the $1.3M_{\odot}$ peak \cite{Belczynski:private}.  Thus, population synthesis results support the anticipation that NS binaries may have a narrow range of masses that could be modeled by a Gaussian distribution.

To lowest order, the GW signal depends on the two neutron star masses through the chirp mass, $\mathcal{M}$. We assume that the distribution of individual neutron-star masses is normal, as suggested above.  For $\sigma_{\rm{NS}} \ll \mu_{\rm{NS}}$, this should yield an approximately normal distribution for the chirp mass as well.

We carried out ${\sim}O(10^5)$ iterations, drawing two random variates from a normal distribution (representing the individual neutron-star masses), and then computing $\mathcal{M}$. We varied the mean and width of the underlying distribution within the allowed ranges (Sec.\ \ref{sec:detector}). Binning the $\mathcal{M}$ values, the resulting $\mathcal{M}$ distribution was found to be normal, as expected.

We postulate a simple ansatz for the relationship between the chirp mass distribution parameters and the underlying neutron star mass distribution. If $X_1$ and $X_2$ are two independent random variates drawn from normal distributions,
\begin{align} \label{eq:normal}
X_1\sim N({\mu}_1, {\sigma}_1^2){\quad}&;{\quad}X_2\sim N({\mu}_2, {\sigma}_2^2)\nonumber\\
aX_1+bX_2\sim N(a{\mu}_1&+b{\mu}_2, a^2{\sigma}_1^2+b^2{\sigma}_2^2).
\end{align}
Since the neutron-star mass distribution is symmetric around the mean (and all neutron star masses are ${\sim}O(1M_{\odot})$ with the values spread over a relatively narrow range), then we can assume a characteristic value for the pre-factor in Eq.\ (\ref{eq:chirp}) is the value taken when both masses are equal i.e.\ ${\sim}(0.25)^{3/5}$. The chirp mass distribution should then be approximately normal
\begin{equation}
{\mathcal{M}}\sim N({\mu}_{c}, \sigma_{c}^2), \nonumber
\end{equation}
{\noindent}with mean and standard deviation
\begin{equation} \label{eq:ansatz}
{\mu}_c\approx 2(0.25)^{3/5}{\mu}_{\rm{NS}},\quad{\sigma}_c\approx {\sqrt{2}}(0.25)^{3/5}{\sigma}_{\rm{NS}}.
\end{equation}
{\noindent}where ${\mu}_{\rm{NS}}$ and ${\sigma}_{\rm{NS}}$ are the mean and standard deviation of the underlying neutron-star mass distribution, respectively.

The accuracy of such an ansatz depends upon the size of mass asymmetries which could arise in a DNS binary system. We investigated the percentage offset between the \textit{actual} distribution parameters (deduced from least-squares fitting to the sample number-density distribution) and the ansatz parameters, for a few values of ${\mu}_{\rm{NS}}$ and ${\sigma}_{\rm{NS}}$. The largest offset of the ansatz parameters from the true chirp mass distribution was on the order of a few percent (${\sim}2.5{\%}$ for $\mu_c$, and ${\sim}3.5{\%}$ for $\sigma_c$ when $\mu_{\rm{NS}}=1.0M_{\odot}$, $\sigma_{\rm{NS}}=0.3M_{\odot}$), and the agreement improved with a narrower underlying neutron star mass distribution. For the case of ${\sigma}_{\rm{NS}}\sim 0.05M_{\odot}$, the agreement was ${\sim}0.1\%$ for ${\mu}_c$ and $<0.1{\%}$ for ${\sigma}_c$. In the case of ${\sigma}_{\rm{NS}}\sim 0.15M_{\odot}$, the agreement was within a percent for both parameters. The sign of these offsets indicates that ${\mu}_c^{\rm{true}}<{\mu}_c^{\rm{model}}$ and ${\sigma}_c^{\rm{true}}>{\sigma}_c^{\rm{model}}$.

Given that the literature indicates an underlying neutron-star mass distribution in DNS systems with ${\sigma}_{\rm{NS}}\lesssim 0.15M_{\odot}$, we anticipate that Eq.\ (\ref{eq:ansatz}) will be appropriate for generating data sets and we use this in the ensuing analysis. The assumption throughout is that for the volume of the Universe probed by our global network, the neutron-star mass distribution does not change.  

The observed data will tell us how wide the intrinsic chirp mass distribution is in reality.  If it is wider than anticipated, we may not be able to measure $H_0$ as precisely as we find here, but we will know this from the observations.  In principle, we could still be systematically biased if the mass distribution turned out to be significantly non-Gaussian, since we are assuming a Gaussian model.  However, it will be fairly obvious if the mass distribution is significantly non-Gaussian (e.g., has a non-negligible secondary peak around $1.8\ M_\odot$), since redshift could only introduce a ${\sim}10\%$ spread in the very precise redshifted chirp mass measurements for detectors that are sensitive to $z\sim0.1$.  In such a case, we would not attempt to fit the data to a Gaussian model for the intrinsic chirp mass distribution.

\subsection{DNS binary merger rate density, ${\dot{n}}(z)$}

We assume that merging DNS systems are distributed homogeneously and isotropically. The total number of these systems that will be detected by the global network depends on the intrinsic rate of coalescing binary systems \textit{per comoving volume}. We require some knowledge of this in order to generate our mock data sets. Any sort of redshift evolution of this quantity (as a result of star-formation rate evolution etc.) can be factorised out \citep{Finn96}, such that
\begin{equation}
\dot{n}(t)\equiv{\frac{d^2N}{dt_{\rm e}dV_{c}}}\equiv\dot{n}(z)={{\dot{n}}_0}{\xi(z)},
\end{equation}
{\noindent}where $N$ is the number of coalescing systems, $t_{\rm e}$ is proper time, $V_c$ is comoving volume, and ${\dot{n}}_0$ represents the local merger-rate density.

We will consider an evolving merger-rate density, such that,
\begin{equation} \label{eq:density-evolve}
{\xi(z)} = 1+{\alpha}z = 1+2z,\quad\quad\text{for }z\leq1,
\end{equation}
{\noindent}which is motivated by a piecewise linear fit \citep{cutler-harms} to the merger-rate evolution deduced from the UV-luminosity-inferred star-formation-rate history \citep{schneider2001}. 

The appropriate value for ${\dot{n}}_0$ is discussed in detail in \cite{abadie-rate2010}. In that paper, the authors review the range of values quoted in the literature for compact binary coalescence rates, i.e.\ not only NS-NS mergers but NS-BH and BH-BH. The binary coalescence rates are quoted per Milky Way Equivalent Galaxy (MWEG) and per $L_{10}$ ($10^{10}$ times the Solar blue-light luminosity, $L_{B,{\odot}}$), as well as per unit comoving volume. In each case, the rates are characterized by four values, a ``low,'' ``realistic,'' ``high'' and ``maximum'' rate, which cover the full range of published estimates. 

\begin{table} 
\caption{\label{tab:rate-data}A compilation of NS-NS merger rate densities in various forms from Tables II, III and IV in \cite{abadie-rate2010}. The first column gives the units. The second, third and fourth columns denote the plausible pessimistic, likely, and plausible optimistic merger rates extrapolated from the observed sample of Galactic binary neutron stars \cite{kalogera2004}. The fifth column denotes the upper rate limit deduced from the rate of Type Ib/Ic supernovae \citep{kim2006}.}
\begin{ruledtabular}
\begin{tabular}{l c c c c}
Source & $R_{\rm{low}}$ & $R_{\rm{re}}$ & $R_{\rm{high}}$ & $R_{\rm{max}}$ \\
\hline
NS-NS (MWEG$^{-1}$Myr$^{-1}$) & 1 & 100 & 1000 & 4000 \\
NS-NS (L$_{10}^{-1}$Myr$^{-1}$) & 0.6 & 60 & 600 & 2000  \\
NS-NS (Mpc$^{-3}$Myr$^{-1}$) & 0.01 & 1 & 10 & 50  \\
\end{tabular}
\end{ruledtabular}
\end{table}

The values for the NS-NS merger rate given by \cite{abadie-rate2010} are listed in Table \ref{tab:rate-data}. The second row of Table \ref{tab:rate-data} is derived assuming that coalescence rates are proportional to the star-formation rate in nearby spiral galaxies. This star-formation rate is crudely estimated from their blue-luminosity, and the merger-rate density is deduced via the conversion factor of $1.7$ $L_{10}/$MWEG \citep{kalogera2001}. The data in the third row is obtained using the conversion factor of $0.0198$ $L_{10}/$Mpc$^{3}$ \citep{koppa2008}.

To convert from merger-rate densities to \textit{detection rates}, \cite{abadie-rate2010} take the product of the merger-rate density with the volume of a sphere with radius equal to the \textit{volume averaged horizon distance}. The horizon distance is the distance at which an optimally oriented, optimally located binary system of inspiraling $1.4M_{\odot}$ neutron stars is detected with the threshold SNR. This is then averaged over all sky locations and binary orientations.
\begin{equation} \label{eq:abadie-calc}
N_{D}={\dot{n}_{0}}{\times}{\frac{4\pi}{3}}\left({\frac{D_{\rm{horizon}}}{\rm{Mpc}}}\right)^3(2.26)^{-3},
\end{equation}
{\noindent}where the $(1/2.26)$ factor represents the average over all sky locations and binary orientations. 

This gives ${\sim}40$ detection events per year in AdLIGO (using $R_{\rm{re}}$), assuming that $D_{\rm{horizon}}=445$ Mpc and all neutron stars have a mass of $1.4M_{\odot}$. 

\subsection{Cosmological model assumptions}\label{sec:cosmo}
We assume a flat cosmology, ${\Omega}_{k,0}=0$, throughout, for which the luminosity distance as a function of the radial comoving distance is given by
\begin{equation}
D_L(z) = (1+z)D_c(z) = (1+z) D_H {\int_0^z{\frac{dz'}{E(z')}}},
\end{equation}
{\noindent}where $D_H={c}/{H_0}$ (the ``\textit{Hubble length scale}'') and
\begin{equation}
E(z)=\sqrt{{\Omega}_{m,0}(1+z)^3+{\Omega}_{{\Lambda},0}}.
\end{equation}
In such a cosmology, the redshift derivative of the comoving volume is given by
\begin{equation}
{\frac{dV_c}{dz}}=\frac{4{\pi}{D_c(z)}^2D_H}{E(z)}.
\end{equation}
 At low redshifts, we can use an approximate simplified form for the relationship between redshift and luminosity distance.  Using a Taylor expansion of the comoving distance around $z=0$ up to $O(z^2)$, and taking the appropriate positive root, we find $D_c(z)=D_L(z)/(1+z)$ is given by
\begin{align}
D_c(z)&\approx{D_c(z=0)}+z\,{\frac{\partial D_c}{\partial z}}\bigg\vert_{z=0}+{\frac{z^2}{2!}}\,{\frac{\partial^{2} D_c}{\partial z^2}}\bigg\vert_{z=0}+\ldots\nonumber\\
&\approx D_H\left[z-{\frac{3}{4}}{\Omega}_{m,0}z^2\right]
\end{align}
Hence,
\begin{equation}
D_L\approx D_H\left[z+\left(1-{\frac{3}{4}}{\Omega}_{m,0}\right)z^2\right]\nonumber
\end{equation}
Therefore,
\begin{equation} \label{eq:redshift}
z\approx{\frac{1}{2\left(1-{\frac{3}{4}}{\Omega}_{m,0}\right)}}\left[\sqrt{1+\frac{4\left(1-{\frac{3}{4}}{\Omega}_{m,0}\right)D_L}{D_H}}-1\right].
\end{equation}
This approximation is very accurate for the range of parameters investigated ($H_0\in[0,200]$ km s$^{-1}$Mpc$^{-1}$, ${\Omega}_{m,0}\in[0,0.5]$), and for $D_L\lesssim1$ Gpc (which is comfortably beyond the reach of AdLIGO for NS-NS binaries). In this parameter range, the largest offset of this approximation from a full redshift root-finding algorithm is ${\sim}4.6{\%}$, at a luminosity distance of $1$ Gpc.

\subsection{Distribution of detectable DNS systems} \label{sec:distribution-detect}

The two system properties we will use in our analysis are the redshifted chirp mass, $\mathcal{M}_z$, and the luminosity distance, $D_L$. Only systems with an SNR greater than threshold will be detected. Thus, we must include SNR selection effects in the calculation for the number of detections. We can write down the distribution of the number of events per year with $\mathcal{M}$, $z$ and $\Theta$ \citep{Finn96,oshaughnessy2010},
\begin{equation} \label{eq:data-generate}
\frac{d^4N}{dtd{\Theta}dzd\mathcal{M}}={\frac{dV_c}{dz}}{\frac{\dot{n}(z)}{(1+z)}}{\mathcal{P}}({\mathcal{M}}){\mathcal{P}}_{\Theta}(\Theta),
\end{equation}
{\noindent}where $t$ is the time measured in the observer's frame, such that the $1/(1+z)$ factor accounts for the redshifting of the merger rate \citep{oshaughnessy2010}.

We convert this to a distribution in $\mathcal{M}_z$, $D_L$ and $\rho$ using,
\begin{equation}
{\frac{d^4N}{dtd{\rho}d{D_L}d\mathcal{M}_z}}=
 \begin{vmatrix}
  {\frac{\partial \mathcal{M}}{\partial {\mathcal{M}}_{z}}} & {\frac{\partial \mathcal{M}}{\partial D_L}} & {\frac{\partial \mathcal{M}}{\partial \rho}}\\
  {\frac{\partial z}{\partial {\mathcal{M}}_{z}}} & {\frac{\partial z}{\partial D_L}} & {\frac{\partial z}{\partial \rho}} \\
  {\frac{\partial \Theta}{\partial {\mathcal{M}}_{z}}} & {\frac{\partial \Theta}{\partial D_L}} & {\frac{\partial \Theta}{\partial \rho}}
 \end{vmatrix}
\times\frac{d^4N}{dtd{\Theta}dzd\mathcal{M}}.
\end{equation}
We use the definitions of the variables in Sec.\ \ref{sec:properties} and \ref{sec:detector} to evaluate the Jacobian matrix determinant. The redshift is only a function of $D_L$ (in a given cosmology); the intrinsic chirp mass, $\mathcal{M}$, is the redshifted chirp mass divided by $(1+z)$ (again the redshift is a function of $D_L$); $\Theta$ is a function of $\mathcal{M}_z$, $D_L$ and $\rho$ according to Eq.\ (\ref{eq:snr}). The (1,3) component $\left({\partial \mathcal{M}}/{\partial \rho}\equiv({\partial \mathcal{M}}/{\partial \rho})\big\vert_{{\mathcal{M}}_{z},D_L}\right)$ is zero because we are differentiating intrinsic chirp mass (a function of redshifted chirp mass and distance) with respect to SNR, but keeping distance and redshifted chirp mass constant. If these variables are held constant then the derivative must be zero. Similar considerations of which variables are held constant in the partial derivatives are used to evaluate the remaining elements of the matrix. Hence, 
\begin{align}
\begin{vmatrix}
  \frac{1}{(1+z)} & -{\frac{\mathcal{M}}{(1+z)}}{\frac{\partial z}{\partial D_L}} & 0 \\
  0 & {\frac{\partial z}{\partial D_L}} & 0 \\
  -{\frac{5}{6}}{\frac{\Theta}{{\mathcal{M}}_z}} & \frac{\Theta}{D_L} & \frac{\Theta}{\rho}
 \end{vmatrix}
&=\frac{1}{(1+z)}\frac{\partial z}{\partial D_L}{\frac{\Theta}{\rho}}.
\end{align}
We note that,
\begin{equation}
{\mathcal{P}}_{\rho}(\rho){\delta}{\rho}={\mathcal{P}}_{\Theta}(\Theta){\delta}{\Theta},\nonumber
\end{equation}
which gives,
\begin{align}{\mathcal{P}}_{\rho}(\rho|{\mathcal{M}}_{z},D_L)=&{\mathcal{P}}_{\Theta}(\Theta){\frac{\partial \Theta}{\partial \rho}}\bigg\vert_{{\mathcal{M}}_{z},D_L}={\mathcal{P}}_{\Theta}(\Theta){\frac{\Theta}{\rho}}\nonumber\\
=&{\mathcal{P}}_{\Theta}\left[{\frac{\rho}{8}}{\frac{D_L}{r_{0}}}\left({\frac{1.2M_{\odot}}{{\mathcal{M}}_{z}}}\right)^{5/6}\right]\nonumber\\
&\times{\frac{D_L}{8r_{0}}}\left({\frac{1.2M_{\odot}}{{\mathcal{M}}_{z}}}\right)^{5/6},
\end{align}
{\noindent}such that we finally obtain,
\begin{align} \label{eq:distribution}
{\frac{d^4N}{dtd{\rho}d{D_L}d\mathcal{M}_z}}=&\frac{1}{(1+z)}\frac{\partial z}{\partial D_L}{\frac{dV_c}{dz}}{\frac{\dot{n}(z)}{(1+z)}}\nonumber\\
&\times{\mathcal{P}}({\mathcal{M}}|z)\times\underbrace{{\mathcal{P}}_{\Theta}(\Theta){\frac{\Theta}{\rho}}}_{{\mathcal{P}}_{\rho}({\rho}|{\mathcal{M}}_{z}, D_L)}\nonumber\\
=&{\frac{4{\pi}{D_c(z)}^2D_H}{D_c(z)E(z)+D_H(1+z)}}{\frac{{\dot{n}}(z)}{(1+z)^2}}\nonumber\\
&\times{\mathcal{P}}\left({\frac{{\mathcal{M}}_{z}}{1+z}}{\bigg\vert}{D_L}\right){\mathcal{P}}_{\rho}({\rho}|{\mathcal{M}}_{z}, D_L).
\end{align}
We may not necessarily care about the specific SNR of a detection; rather only that a system with ${\mathcal{M}}_{z}$ and $D_L$ has SNR above threshold (and is thus detectable). Fortunately the SNR only enters Eq.\ (\ref{eq:distribution}) through ${\mathcal{P}}_{\rho}({\rho}|{\mathcal{M}}_{z}, D_L)$, such that we can simply integrate over this term and apply Eq.\ (\ref{eq:ctheta}), 
\begin{align}
{\int_{\rho_{0}}^{\infty}}{\mathcal{P}}_{\rho}(\rho|{\mathcal{M}}_{z},D_L)d{\rho}=&{\int_{x}^{\infty}}{\mathcal{P}}_{\Theta}(\Theta)d{\Theta}\equiv C_{\Theta}(x),\nonumber\\
&\text{where,}\quad x={\frac{\rho_0}{8}}{\frac{D_L}{r_{0}}}\left({\frac{1.2M_{\odot}}{{\mathcal{M}}_{z}}}\right)^{5/6}.
\end{align}
In this case, Eq.\ (\ref{eq:distribution}) is modified to give,
\begin{align} \label{eq:rprob}
&{\frac{d^3N}{dtd{D_L}d\mathcal{M}_z}}{\bigg\vert}_{{\rho}>{\rho}_0}\nonumber\\
=&{\frac{4{\pi}{D_c(z)}^2D_H}{D_c(z)E(z)+D_H(1+z)}}{\frac{{\dot{n}}(z)}{(1+z)^2}}\nonumber\\
&\times{\mathcal{P}}\left({\frac{{\mathcal{M}}_{z}}{1+z}}{\bigg\vert}{D_L}\right)C_{\Theta}\left[{\frac{\rho_0}{8}}{\frac{D_L}{r_{0}}}\left({\frac{1.2M_{\odot}}{{\mathcal{M}}_{z}}}\right)^{5/6}\right].
\end{align}
To calculate the number of detected systems (given a set of cosmological and NS mass distribution parameters, $\overrightarrow\mu$) we integrate over this distribution, which is equivalent to integrating over the distribution of events with redshift and chirp mass, i.e.\ $N_{\mu}=T\times{\int_0^{\infty}}{\int_0^{\infty}}\left(\frac{d^3N}{dtdzd\mathcal{M}}\right)dzd{\mathcal{M}}$,
{\noindent}where $T$ is the duration of the observation run.

\subsection{Creating mock catalogs of DNS binary inspiraling systems} \label{sec:ref-model}
\begin{table}
\caption{A summary of the WMAP 7-year observations. The data from Column 1 is from Table 3 of \cite{wmap_only}, containing parameters derived from fitting models to WMAP data only. Column 2 contains the derived parameters from Table 8 of \cite{wmap_plus}, where the values result from a six-parameter flat $\Lambda$CDM model fit to WMAP+BAO+SNe data.\label{tab:cosmo-data}}
\begin{ruledtabular}
\begin{tabular}{c c c c c}
Parameter & WMAP only & WMAP+BAO+SNe\\
\hline
{$H_0$ / (km\ s$^{-1}$Mpc$^{-1}$)} & {$71.0\pm2.5$} & {$70.4_{-1.4}^{+1.3}$}\\
{${\Omega}_{b,0}$} & {$0.0449\pm0.0028$} & {$0.0456\pm0.0016$}\\
{${\Omega}_{c,0}$} & {$0.222\pm0.026$} & {$0.227\pm0.014$}\\
{${\Omega}_{\Lambda,0}$} & {$0.734\pm0.029$} & {$0.728_{-0.016}^{+0.015}$}\\
\end{tabular}
\end{ruledtabular}
\end{table}
The model parameter space we investigate is the $5$D space of $[H_0,{\mu}_{\rm{NS}},{\sigma}_{\rm{NS}},{\Omega}_{m,0},{\alpha}]$ with a flat cosmology assumed. To generate a catalog of events, we choose a set of reference parameters, motivated by previous analysis in the literature. The seven-year WMAP observations gave the cosmological parameters in Table \ref{tab:cosmo-data}. For our reference cosmology, we adopt $H_0=70.4${\hspace{1mm}} km\ s$^{-1}$Mpc$^{-1}$, ${\Omega}_{m,0}=0.27$ and ${\Omega}_{\Lambda,0}=0.73$. The parameters of the neutron-star mass distribution were discussed earlier, but as reference we use ${\mu}_{\rm{NS}}=1.35M_{\odot}$ and ${\sigma}_{\rm{NS}}=0.06M_{\odot}$. The merger-rate density was also discussed earlier, and we take ${\alpha}=2.0$ and ${\dot{n}}_0=10^{-6}$ Mpc$^{-3}$yr$^{-1}$ as the reference. Later, we will investigate how the results change if the width of the NS mass distribution is as large as $0.13M_{\odot}$, as indicated by the predictive density estimate of \cite{kiziltan2010}.

These reference parameters are used to calculate an expected number of events,\footnote{The observation time, $T$, is assumed to be $1$ year (but the expected number of detections simply scales linearly with time) and a network acting as a super-interferometer with $r_{0,\rm{net}}\simeq176$ Mpc is also assumed.} and the number of observed events is drawn from a Poisson distribution (assuming each binary system is independent of all others) with that mean. Monte-Carlo acceptance/rejection sampling is used to draw random redshifts and chirp masses from the distribution in Eq.\ (\ref{eq:data-generate}) for each of the $N_o$ events. The $D_L$ and ${\mathcal{M}}_z$ are then computed from the sampled  ${\mathcal{M}}$ and $z$.

With a reference rate of ${\dot{n}}_0=10^{-6}$ Mpc$^{-3}$yr$^{-1}$ and a constant merger-rate density, we estimate that there should be ${\sim}90$ yr$^{-1}$ detections, whilst taking into account merger-rate evolution using Eq.\ (\ref{eq:density-evolve}) boosts this to ${\sim}100$ yr$^{-1}$. These numbers are for a network SNR threshold of $8$. 
If we ignore merger-rate evolution and raise the SNR threshold to $10$ (to represent an AdVirgo-HHL network for which the coincident detection rate is roughly halved relative to the HHL-only network) we get ${\sim}45$ events in $1$ year, which compares well to the $40$ events found in \cite{abadie-rate2010}.

\section{Analysis methodology} \label{sec:analysis-method}

We will use Bayesian analysis techniques to simultaneously compute posterior distribution functions on the mean and standard deviation of the intrinsic NS mass distribution (in DNS systems) and the cosmological parameters given a catalog of simulated sources with measured redshifted chirp masses and luminosity distances.

\subsection{Bayesian analysis using  Markov Chain Monte Carlo techniques} \label{sec:bayes}

Bayes' theorem states that the inferred \textit{posterior} probability distribution of the parameters $\overrightarrow{{\mu}}$ based on a hypothesis model $\mathcal{H}$, and given data $D$ is given by
\begin{equation} \label{eq:bayes-theorem}
p(\overrightarrow{{\mu}}|D,\mathcal{H}) = \frac{{\mathcal{L}}(D|\overrightarrow{{\mu}},\mathcal{H})\pi(\overrightarrow{{\mu}}|\mathcal{H})}{p(D|\mathcal{H})},
\end{equation}
where ${\mathcal{L}}(D|\overrightarrow{{\mu}},\mathcal{H})$ is the \textit{likelihood} (the probability of measuring the data, given a model with parameters $\overrightarrow{{\mu}}$), $\pi(\overrightarrow{{\mu}}|\mathcal{H})$ is the \textit{prior} (any constraints already existing on the model parameters) and finally $p(D|\mathcal{H})$ is the \textit{evidence} (this is important in model selection, but in the subsequent analysis in this paper can be ignored as a normalization constant).

In this analysis, the data in Eq.\ (\ref{eq:bayes-theorem}) is not from a single source, but rather from a set of sources, and we want to use it to constrain certain aspects of the source distribution, as well as the background cosmology. The uncertainty arises from the fact that any model cannot predict the exact events we will see, but rather an astrophysical rate of events that gives rise to the observed events.
The probability distribution for the set of events will be discussed in Sec.\ \ref{sec:like}.

To compute the posterior on the model parameters, we use Markov Chain Monte Carlo (MCMC) techniques since they provide an efficient way to explore the full parameter space. An initial point, $\overrightarrow{x_0}$, is drawn from the \textit{prior} distribution and then at each subsequent iteration, $i$, a new point, $\overrightarrow{y}$, is drawn from a \textit{proposal distribution}, $q({\overrightarrow{y}}|{\overrightarrow{x}})$ (uniform in all cases, covering the range of parameter investigation). The Metropolis-Hastings ratio is then evaluated,
\begin{equation}
R=\frac{{\pi}(\overrightarrow{y}){\mathcal{L}}(D|{\overrightarrow{y}},\mathcal{H})q({\overrightarrow{x_i}}|\overrightarrow{y})}{{\pi}(\overrightarrow{x_i}){\mathcal{L}}(D|{\overrightarrow{x_i}},\mathcal{H})q({\overrightarrow{y}}|{\overrightarrow{x_i}})}.
\end{equation}

A random sample is drawn from a uniform distribution, $u\in U[0,1]$, and if $u<R$ the move to the new point is accepted, so that we set ${\overrightarrow{x_{i+1}}}={\overrightarrow{y}}$. If $u>R$, the move is rejected and we set ${\overrightarrow{x_{i+1}}}={\overrightarrow{x_i}}$. If $R>1$ the move is always accepted, however if $R<1$ there is still a chance that the move will be accepted.

The MCMC samples can be used to carry out integrals over the posterior, e.g.\ 
\begin{equation}
\int f(\overrightarrow{x})p(\overrightarrow{x}|D,\mathcal{H})d\overrightarrow{x}={\frac{1}{N}}\displaystyle\sum_{i=1}^Nf(\overrightarrow{x_i}).
\end{equation}
The $1$D marginalized posterior probability distributions in individual model parameters can be obtained by binning the chain samples in that parameter.

\subsection{Modelling the likelihood} \label{sec:like}
\subsubsection{Expressing the likelihood}
We use a theoretical framework similar to that of \cite{gair2010}. The data are assumed to be a catalog of events for which redshifted chirp mass, ${\mathcal{M}}_{z}$, and luminosity distance, $ D_L$ have been estimated. These two parameters for the events can be used to probe the underlying cosmology and neutron-star mass distribution. In this analysis, we focus on what we can learn about the Hubble constant, $H_0$, the Gaussian mean of the (DNS system) neutron-star mass distribution, ${\mu}_{\rm{NS}}$, and the Gaussian half-width, ${\sigma}_{\rm{NS}}$. We could also include the present-day matter density, ${\Omega}_{m,0}$, however we expect that this will not be well constrained due to the low luminosity distances of the sources. We could also include the gradient parameter, ${\alpha}$, describing the redshift evolution of the merger-rate density.

The measurement errors were discussed earlier (Sec.\ \ref{sec:properties}) and we will account for these later. For the first analysis we assume that the observable properties of individual binaries are measured exactly.

We consider first a binned analysis. We divide the parameter space of ${\mathcal{M}}_{z}$ and $D_{L}$ into bins, such that the data is the number of events measured in a particular range of redshifted chirp mass \textit{and} luminosity distance.  Each binary system can be modeled as independent of all other systems, so that within a given galaxy we can model the number of inspirals that occur within a certain time as a Poisson process, with DNS binaries merging at a particular rate (e.g., \citep{Oshaughnessy:2008,Kim:2003kkl}).  The mean of the Poisson process will be equal to the model-dependent rate times the observation time, and the actual number of inspirals occurring in the galaxy is a random-variate drawn from the Poisson distribution. 

A bin in the space of system properties may contain events from several galaxies, but these galaxies will behave independently and the number of recorded detections in a given bin will then be a Poisson process, with a mean equal to the model-dependent \textit{expected} number of detections in that bin \citep{gair2010}. The data can be written as a vector of numbers in labelled bins in the 2D space of system properties, i.e.\ $\overrightarrow{n}=(n_1,n_2,{\ldots},n_X)$, where $X$ is the number of bins. Therefore, the likelihood of recording data $D$ under model $\mathcal{H}$ (with model parameters $\overrightarrow{\mu}$) is the product of the individual Poisson probabilities for detecting $n_i$ events in a bin, $i$, where the expected (model-dependent) number of detections is $r_{i}({\overrightarrow{\mu}})$. For the $i^{\rm{th}}$ bin,
\begin{equation}
p(n_i|{\overrightarrow{\mu}},\mathcal{H})=\frac{(r_{i}({\overrightarrow{\mu}}))^{n_i}e^{-r_{i}({\overrightarrow{\mu}})}}{n_i!},
\end{equation}
and so the likelihood of the cataloged detections is,
\begin{equation} \label{eq:bin-prob}
{\mathcal{L}}({\overrightarrow{n}}|{\overrightarrow{\mu}},\mathcal{H})={\displaystyle\prod_{i=1}^X}\frac{(r_{i}({\overrightarrow{\mu}}))^{n_i}e^{-r_{i}({\overrightarrow{\mu}})}}{n_i!}.
\end{equation}
In this work, we take the continuum limit of Eq.\ (\ref{eq:bin-prob}). In this case, the number of events in each infinitesimal bin is either $0$ or $1$. Every infinitesimal bin contributes a factor of $e^{-r_{i}({\overrightarrow{\mu}})}$, whilst the remaining terms in Eq.\ (\ref{eq:bin-prob}) evaluate to $1$ for empty bins, and $r_{i}({\overrightarrow{\mu}})$ for full bins. The product of the exponential factors gives $e^{-N_{\mu}}$, where $N_{\mu}$ is the number of DNS inspiral detections predicted by the model, with parameters $\overrightarrow{\mu}$. The continuum likelihood of a catalog of discrete events is therefore
\begin{equation} \label{eq:cont-prob}
{\mathcal{L}}(\overrightarrow{{\overrightarrow{\Lambda}}}|{\overrightarrow{\mu}},\mathcal{H})={e^{-N_{\mu}}{\displaystyle\prod_{i=1}^{N_o}}}r({\overrightarrow{{\lambda}_i}}|{\overrightarrow{\mu}}),
\end{equation}
{\noindent}where ${\overrightarrow{{\overrightarrow{\Lambda}}}}={\{}{\overrightarrow{{\lambda}_1}},{\overrightarrow{{\lambda}_2}},{\ldots},{\overrightarrow{{\lambda}_{N_o}}}{\}}$ is the vector of measured system properties, with ${\overrightarrow{{\lambda}_i}}=({\mathcal{M}}_{z}, D_L)_{i}$ for system $i$, and $N_o$ is the number of detected systems. Finally, $r({\overrightarrow{{\lambda}_i}}|{\overrightarrow{\mu}})$ is the rate of events with properties ${\mathcal{M}}_{z}$ and $D_L$, evaluated for the $i^{\rm{th}}$ detection under model parameters $\overrightarrow{\mu}$, which is given by Eq.\ (\ref{eq:rprob}).

\subsubsection{Marginalizing over ${\dot{n}}_0$}

We may also modify the likelihood calculation to marginalize over the poorly constrained merger-rate density, ${\dot{n}}_0$.\footnote{A similar technique was used in \cite{sesana2010}, where the total number of events predicted by the model is marginalized over.} This quantity is so poorly known (see Table \ref{tab:rate-data}), that it is preferable to use a new statistic that does not rely on the local merger-rate density, by integrating the likelihood given in Eq.\ (\ref{eq:cont-prob}) over this quantity,
\begin{equation}
\tilde{\mathcal{L}}(\overrightarrow{{\overrightarrow{\Lambda}}}|{\overrightarrow{\mu}},\mathcal{H})={\int_0^{\infty}}{\mathcal{L}}(\overrightarrow{{\overrightarrow{\Lambda}}}|{\overrightarrow{\mu}},\mathcal{H})d{\dot{n}}_0.
\end{equation}
The expected number of detections described in Sec.\ \ref{sec:distribution-detect} can be expressed as,
\begin{equation}
N_{\mu}={\dot{n}}_0\times\int\int{\mathcal{I}}{\hspace{1.0mm}}d{\mathcal{M}}_{z}dD_L,\nonumber
\end{equation}
{\noindent}where,
\begin{align}
\mathcal{I} =& {\frac{4{\pi}{D_c(z)}^2D_H}{D_c(z)E(z)+D_H(1+z)}}{\frac{1+{\alpha}z}{(1+z)^2}}\nonumber\\
&\times{\mathcal{P}}\left({\frac{{\mathcal{M}}_{z}}{1+z}}{\bigg\vert}{D_L}\right)C_{\Theta}\left[{\frac{\rho_0}{8}}{\frac{D_L}{r_{0}}}\left({\frac{1.2M_{\odot}}{{\mathcal{M}}_{z}}}\right)^{5/6}\right],
\end{align}
{\noindent}and,
\begin{equation}
r({\overrightarrow{{\lambda}_i}}|{\overrightarrow{\mu}})={\dot{n}}_0\times{\mathcal{I}_i},
\end{equation}
{\noindent}where $\mathcal{I}_i$ is the integrand evaluated for the $i^{\rm{th}}$ system's properties. Thus,
\begin{align}
\tilde{\mathcal{L}}(\overrightarrow{{\overrightarrow{\Lambda}}}|{\overrightarrow{\mu}},\mathcal{H})=&\int_0^{\infty}\left[\exp{\left(-{\dot{n}}_0\times\int\int{\mathcal{I}}{\hspace{1.0mm}}d{\mathcal{M}}_{z}dD_L\right)}\right.\nonumber\\
&\times\left.\left({\displaystyle\prod_{i=1}^{N_o}}{\dot{n}}_0\times{\mathcal{I}_i}\right)\right]d{\dot{n}}_0\nonumber\\
=&\left(\int_0^{\infty}{{\dot{n}}_0}^{N_o}\times\exp{\left(-{\dot{n}}_0\,\int\int{\mathcal{I}}d{\mathcal{M}}_{z}dD_L\right)}\right.\nonumber\\
&\quad d{\dot{n}}_0\bigg)\times\displaystyle\prod_{i=1}^{N_o}{\mathcal{I}_i}.
\end{align}
The integral, $\int\int{\mathcal{I}}{\hspace{1.0mm}}d{\mathcal{M}}_{z}dD_L$, depends on the underlying model parameters, $\overrightarrow{\mu}$, through $\mathcal{I}$, but it does not depend on ${\dot{n}}_0$. Therefore, defining
\begin{equation}
{\gamma}={\dot{n}}_0\times\int\int\mathcal{I}{\hspace{1.0mm}}d{\mathcal{M}}_{z}dD_L={\dot{n}}_0\times{\delta}.\nonumber
\end{equation}
We note that ${\dot{n}}_0\in[0,{\infty}]$, hence $\gamma\in[0,{\infty}]$. Therefore,
\begin{align} \label{eq:new-stat}
\tilde{\mathcal{L}}(\overrightarrow{{\overrightarrow{\Lambda}}}|{\overrightarrow{\mu}},\mathcal{H})&=\left(\int_0^{\infty}\left({\frac{\gamma}{\delta}}\right)^{N_o}\times e^{-{\gamma}}{\hspace{1.5mm}}{\frac{d{\gamma}}{\delta}}\right)\displaystyle\prod_{i=1}^{N_o}{\mathcal{I}_i}\nonumber\\
&=\underbrace{\left({\int_0^{\infty}}{\gamma}^{N_o}e^{-{\gamma}}d{\gamma}\right)}_{\textbf{independent of $\overrightarrow{\mu}$}}\times{\delta}^{-(N_o+1)}\displaystyle\prod_{i=1}^{N_o}{\mathcal{I}_i}.
\end{align}
We will verify in the analysis that this new likelihood produces results completely consistent with the case where exact knowledge of the merger-rate density is assumed. We note that we did not include a prior on $\dot{n}_0$ in the above, which is equivalent to using a flat prior for $\dot{n}_0 \in [0,\infty]$. This reflects our current lack of knowledge of the intrinsic merger rate, although such a prior is not normalizable. We could implement a normalizable prior by adding a cut-off, but this cut-off should be set sufficiently high that it will not influence the posterior and therefore the result will be equivalent to the above.

\subsection{Calculating the posterior probability}

The likelihood statistic $\tilde{\mathcal{L}}$ is used to marginalize over the poorly constrained local merger-rate density. We use a weakly informative prior on the model parameters, so that it doesn't prejudice our analysis. As a prior on ${\mu}_{\rm{NS}}$ we take a normal distribution with parameters ${\mu}=1.35M_{\odot}$, ${\sigma}=0.13M_{\odot}$. This is motivated by the posterior predictive density estimate for a neutron star in a DNS binary system given in \cite{kiziltan2010}. We take a prior on ${\alpha}$ that is a normal distribution, centred at $2.0$ with a ${\sigma}$ of $0.5$. Uniform priors were used for the other parameters. 

We made sure that the size of the sampled parameter space was large enough to fully sample the posterior distribution, so that we could investigate how well gravitational-wave observations alone could constrain the cosmology and neutron-star mass distribution. The parameter ranges were $H_0\in[0,200]$ km s$^{-1}$Mpc$^{-1}$, ${\Omega}_{m,0}\in[0,0.5]$, ${\mu}_{\rm{NS}}\in[1.0,1.5]M_{\odot}$, ${\sigma}_{\rm{NS}}\in[0,0.3]M_{\odot}$ and ${\alpha}\in[0.0,5.0]$.

To calculate $\tilde{\mathcal{L}}$ for a given point in the model parameter space we must compute the number of detections predicted by those model parameters ($N_{\mu}$), and we need to calculate $z(D_L)$ for that model so that $\mathcal{M}$ can be evaluated. For the sake of computational efficiency, some approximations are used. We have verified that our results are insensitive to these approximations.  Our approximation for $z(D_L)$ was described in Sec.\ \ref{sec:cosmo}. We also used an analytic ansatz to calculate the model-dependent expected number of detections, based on factorizing the contributions from different model parameters. 
The agreement between this ansatz and the full integrated model number is excellent, with the biggest discrepancy being $\lesssim3{\%}$ of the true value. This allows a direct calculation of $N_{\mu}$ without a multi-dimensional integration for each point in parameter space.   


\section{Results $\&$ Analysis} \label{sec:results}

For our first analysis, we will assume that $\mathcal{M}_z$ and $D_L$ for each individual merger are measured perfectly by our observations, so that the $\mathcal{M}_z$ and $D_L$ recorded for the events are the true values. This represents the best case of what we could learn from GW observations. Later, we will consider how the accuracy of the reconstructed model parameters is affected by including measurement errors on the recorded event properties. 
\subsection{Posterior recovery} 
\begin{figure*}
    \subfloat[]{\label{fig:hubble_post}\incgraph{0.35}{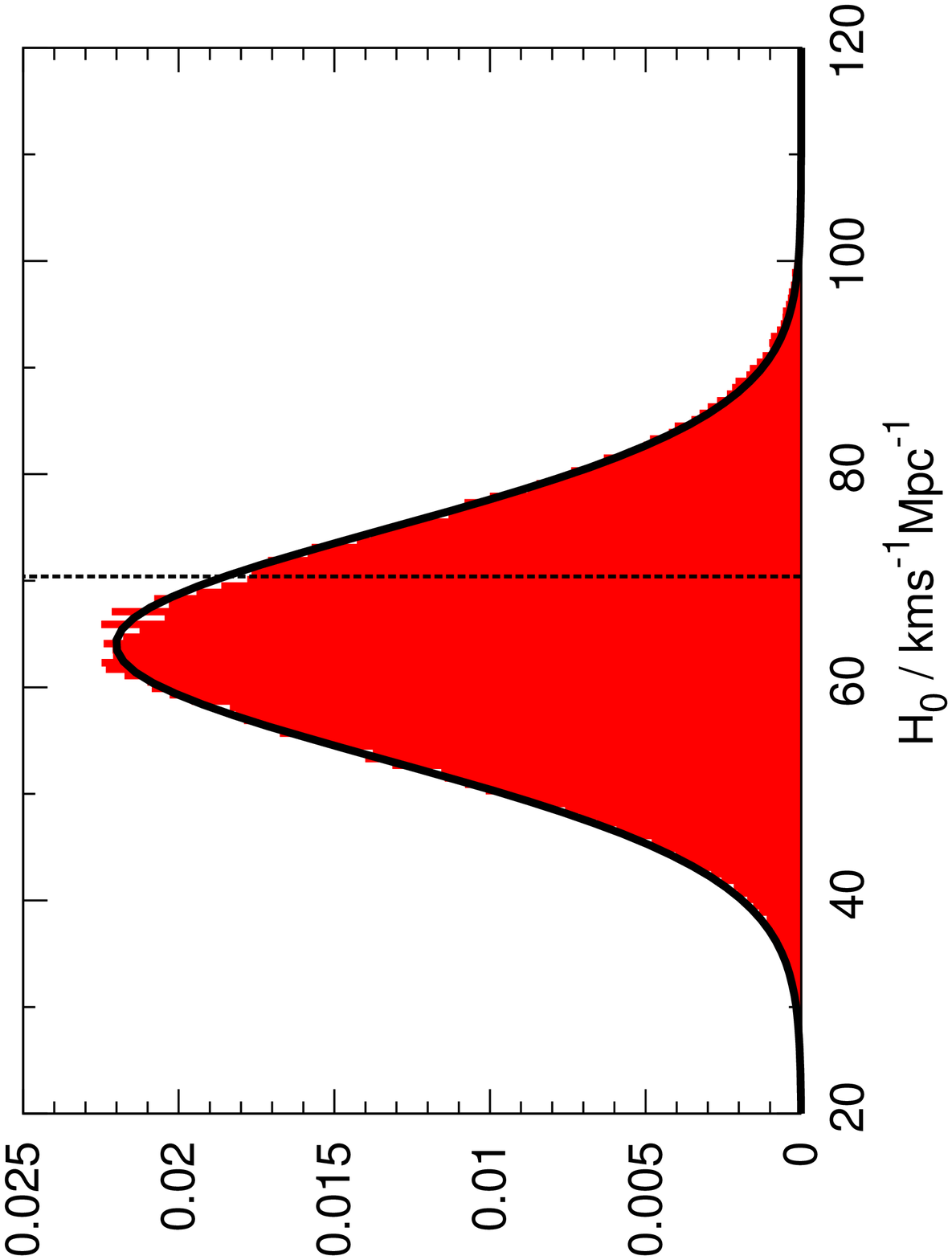}} \hspace{-7mm} 
  \subfloat[]{\label{fig:sigma_post}\incgraph{0.35}{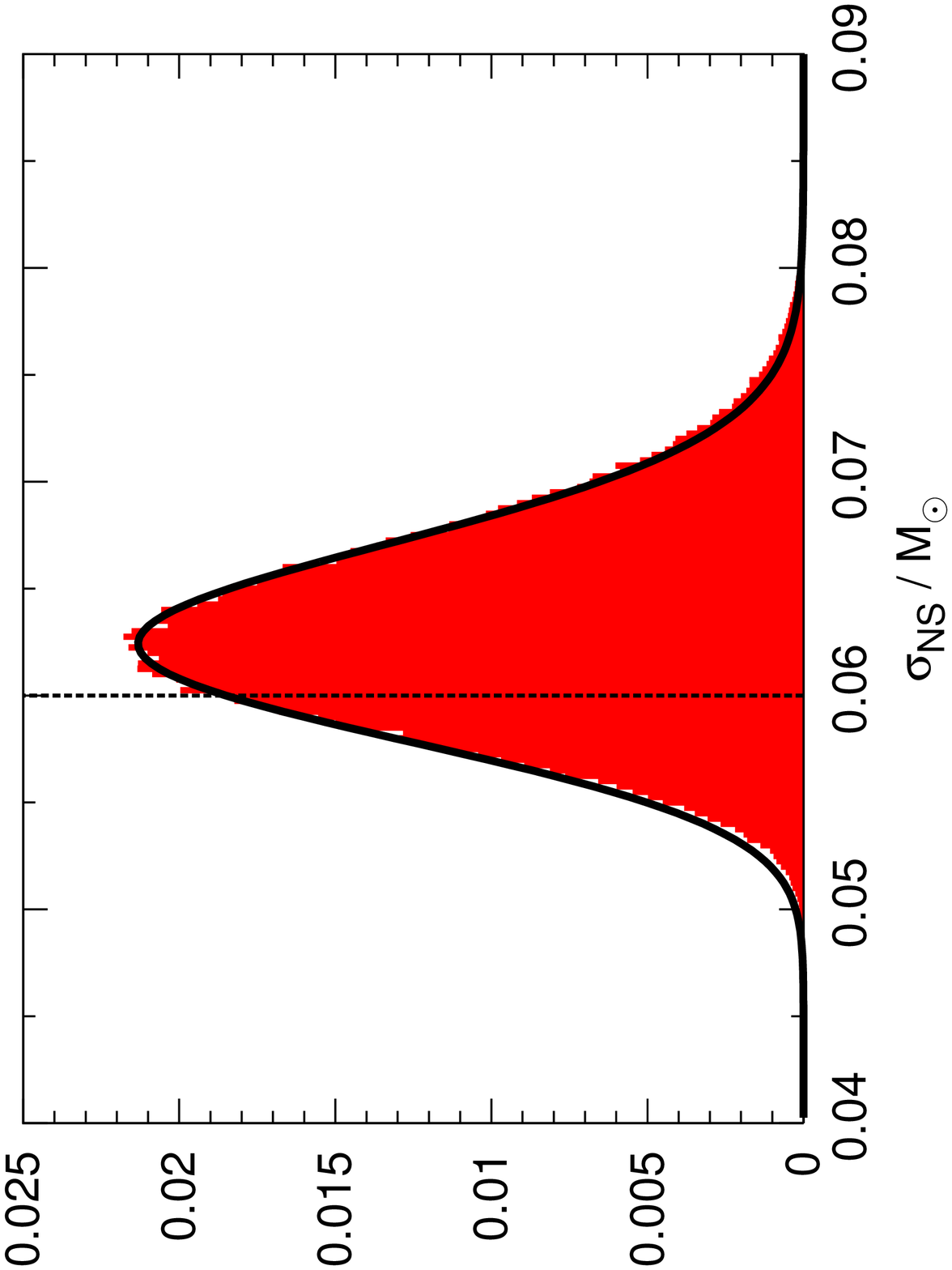}} \hspace{-7mm}
  \subfloat[]{\label{fig:mu_post}\incgraph{0.35}{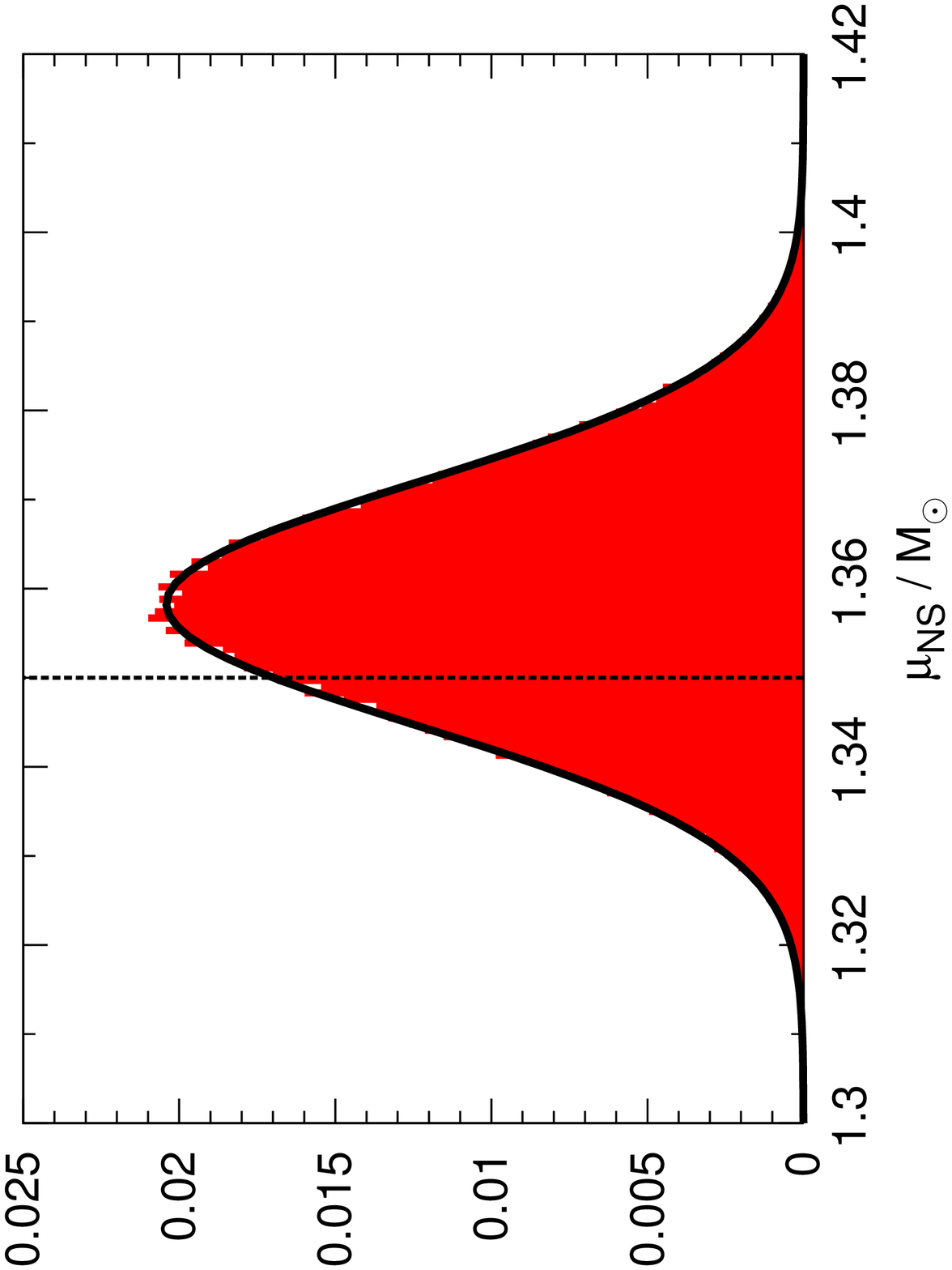}}
  \caption{\label{fig:1d_posteriors}Recovered $1$D posterior distributions for $H_0$ (left), ${\sigma}_{\rm{NS}}$ (centre) and ${\mu}_{\rm{NS}}$ (right), computed for one realization. The black lines represent best-fit Gaussian distributions to $H_0$, ${\ln}({\sigma}_{\rm{NS}})$ and ${\mu}_{\rm{NS}}$, which were obtained via a least-squares fitting procedure. The vertical lines indicate the values of these parameters used to generate the data set.}
\end{figure*}
\begin{figure}
\incgraph{0.45}{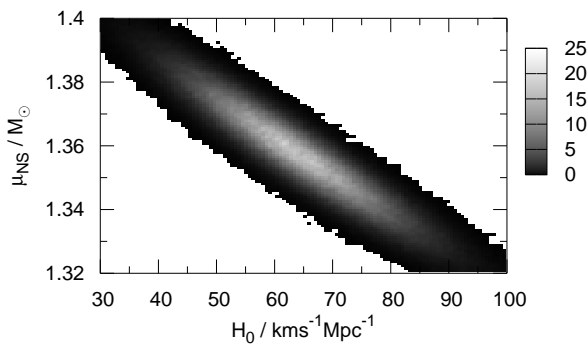}
  \caption{\label{fig:2d_posterior}Recovered $2$D posterior distribution in $H_0$ and ${\mu}_{\rm{NS}}$ space, showing a correlation between these recovered parameters. The model parameter values used to generate the data are the reference values. There appears to be negligible correlation between $\sigma_{\rm{NS}}$ with $H_0$ or $\mu_{\rm{NS}}$.}
\end{figure}
We carried out the analysis discussed in Sec.\ \ref{sec:analysis-method} for a data set calculated from the reference model given in Sec.\ \ref{sec:ref-model}. We found that ${\sim}10^{6}-10^{7}$ samples were necessary for the MCMC analysis to recover the underlying posterior distributions.

We found that ${\Omega}_{m,0}$ and ${\alpha}$ were not constrained by the observations, but their inclusion in the parameter space did not affect our ability to recover the other parameters. For this reason, we kept them in the analysis, but all remaining results will be marginalized over these model parameters. Given the low redshift range that a second-generation network is sensitive to, it is not surprising that the matter-density and merger-rate evolution were not constrained. 

The recovered $1$D posterior distributions in the other parameters are shown in Fig.\ \ref{fig:1d_posteriors} for a typical realization of the set of observed events. We have verified that these marginalized distributions are consistent with those obtained when exact knowledge of the intrinsic ${\dot{n}}_0$ is assumed. We found that the $1$D posterior distributions for $H_0$, ${\ln}({\sigma}_{\rm{NS}})$ and ${\mu}_{\rm{NS}}$ were well fit by Gaussian distributions of the form $A\exp{(-(x-{\mu})^{2}/{2{\sigma}^{2}})}$. These best-fit Gaussians are also shown in the Figure. Although the distributions do not peak at the model parameters used to generate the data, those values are consistent with the mean and width of the recovered distributions. 

In Fig.\ \ref{fig:2d_posterior}, we show the corresponding $2$D posterior distribution in $H_0$ and ${\mu}_{\rm{NS}}$ parameter space. We see that a correlation exists between these parameters. Given a cataloged $D_L$ value, a low value of $H_0$ will imply a low model-dependent redshift. When this redshift is used to compute ${\mathcal{M}}$ from ${\mathcal{M}}_z$, we calculate a large value of the chirp mass, which implies a chirp mass distribution (and hence a neutron-star mass distribution) centered at larger values. ${\sigma}_{\rm{NS}}$ simply encodes the width of the mass distribution around the mean, so on average it should have no effect on $H_0$ and ${\mu}_{\rm{NS}}$ calculations and indeed we found that ${\sigma}_{\rm{NS}}$ showed no correlation with the other model parameters.


It is clear from Fig.\ \ref{fig:1d_posteriors} that the parameters of the Gaussian fits provide a useful way to characterize the recovered distributions. We can then describe the recovered distributions in terms of two best-fit parameters i.e.\ the Gaussian mean, $\mu$, and Gaussian half-width, $\sigma$.

\subsection{Random spread of best-fit parameters}
\subsubsection{No errors in data catalog}
\begin{figure*}             
  \subfloat[]{\incgraph{0.35}{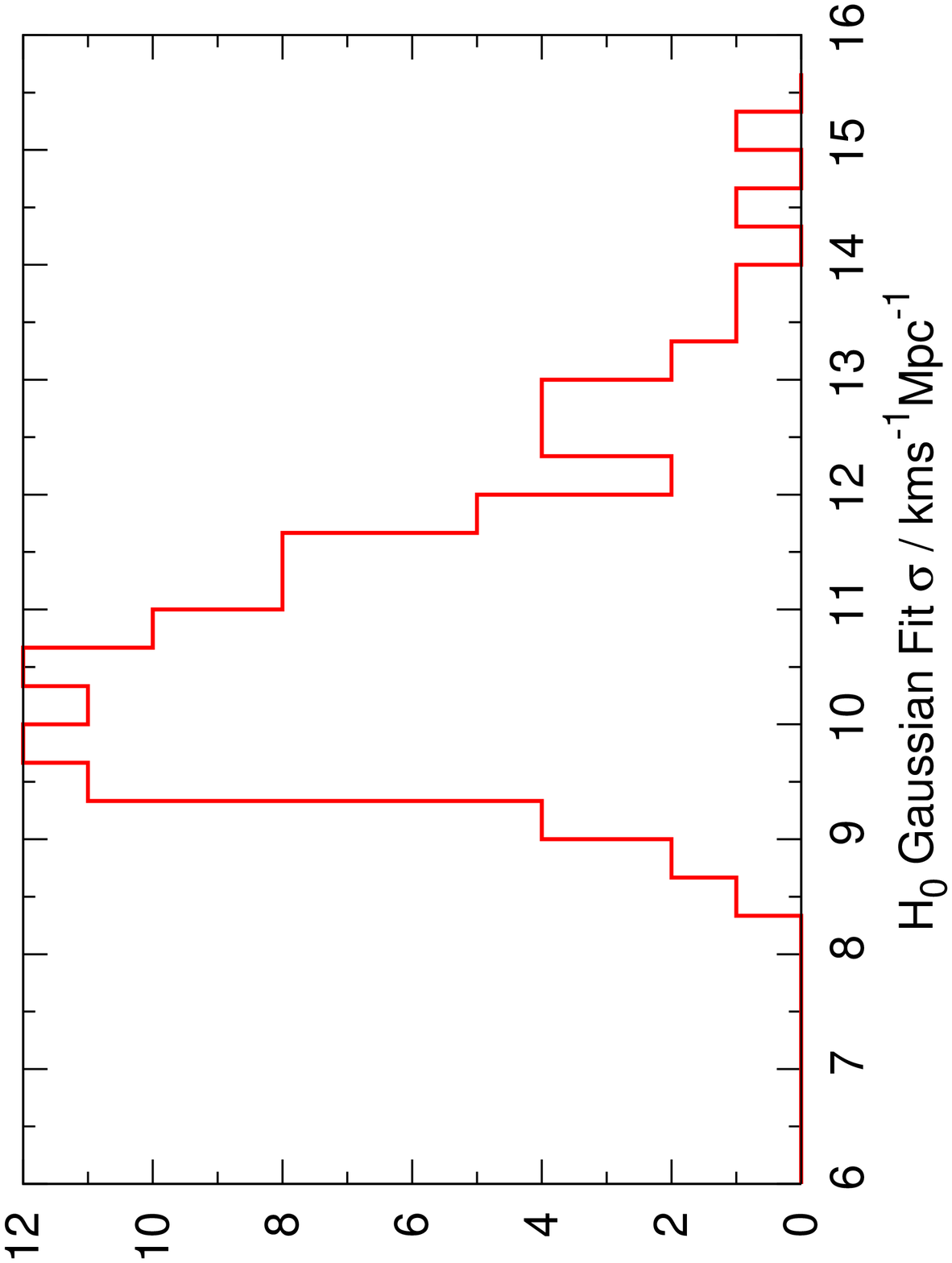}} \hspace{-7mm}               
  \subfloat[]{\incgraph{0.35}{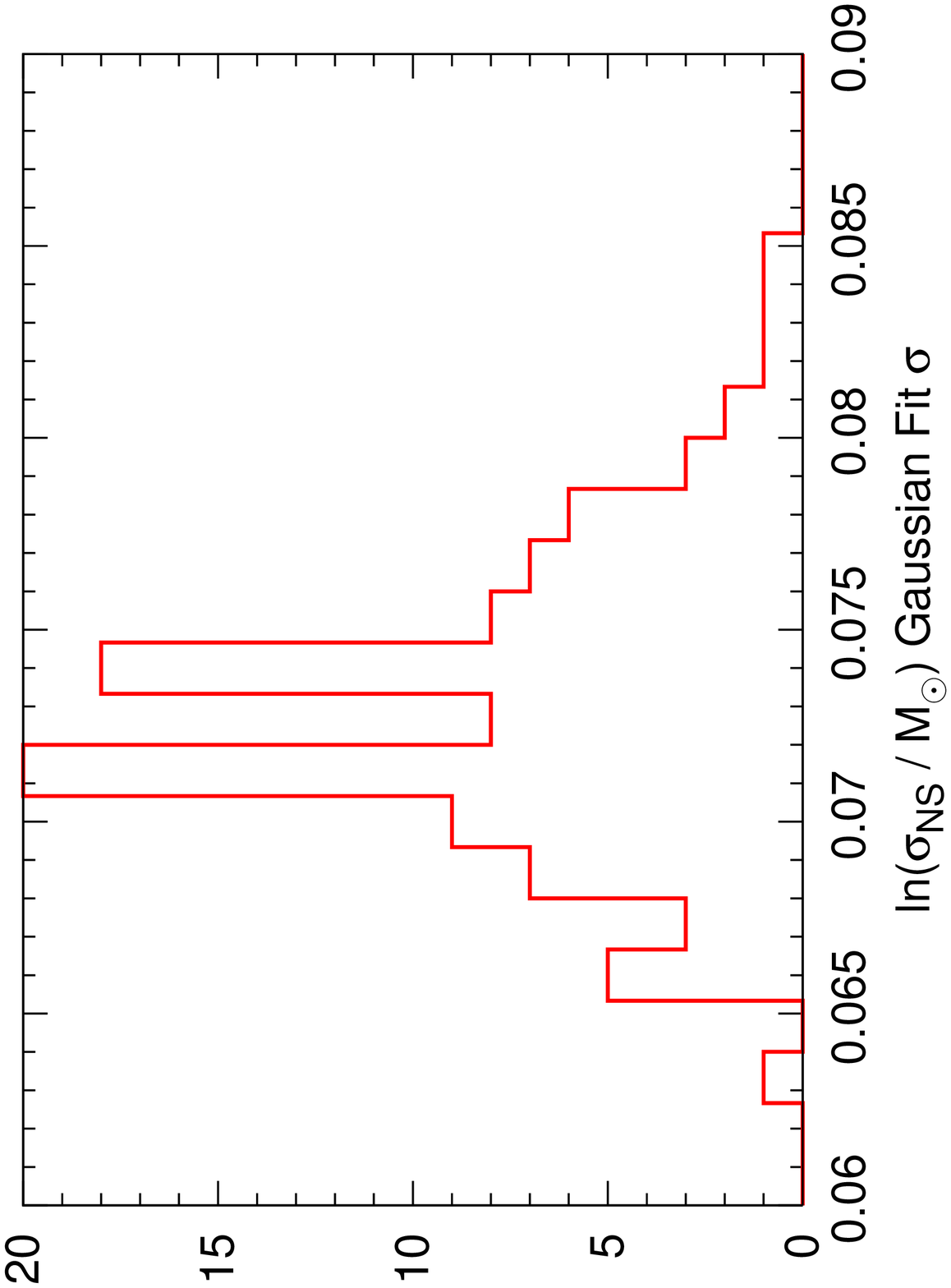}} \hspace{-7mm}                
  \subfloat[]{\incgraph{0.35}{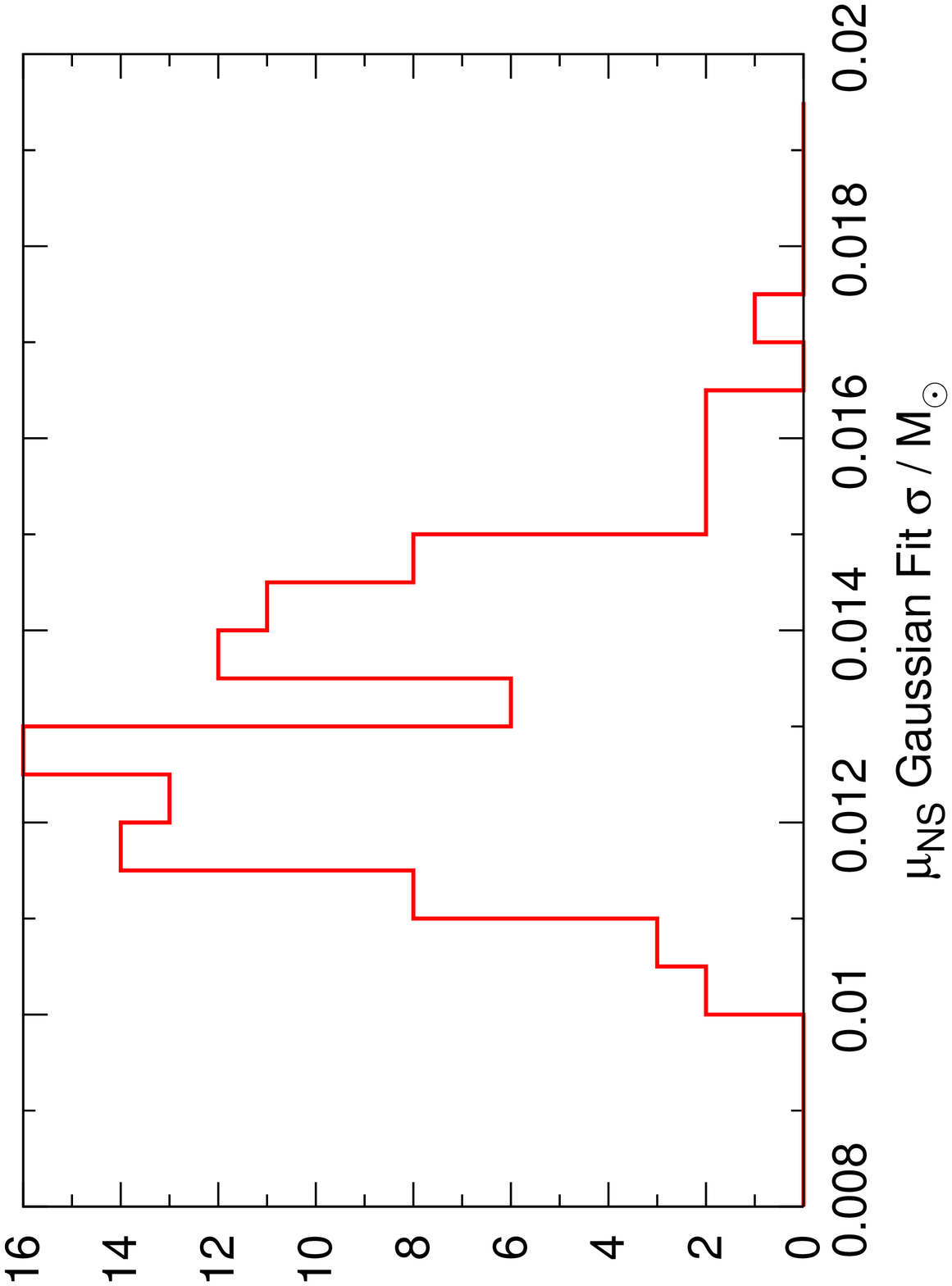}}\\
  \subfloat[]{\incgraph{0.35}{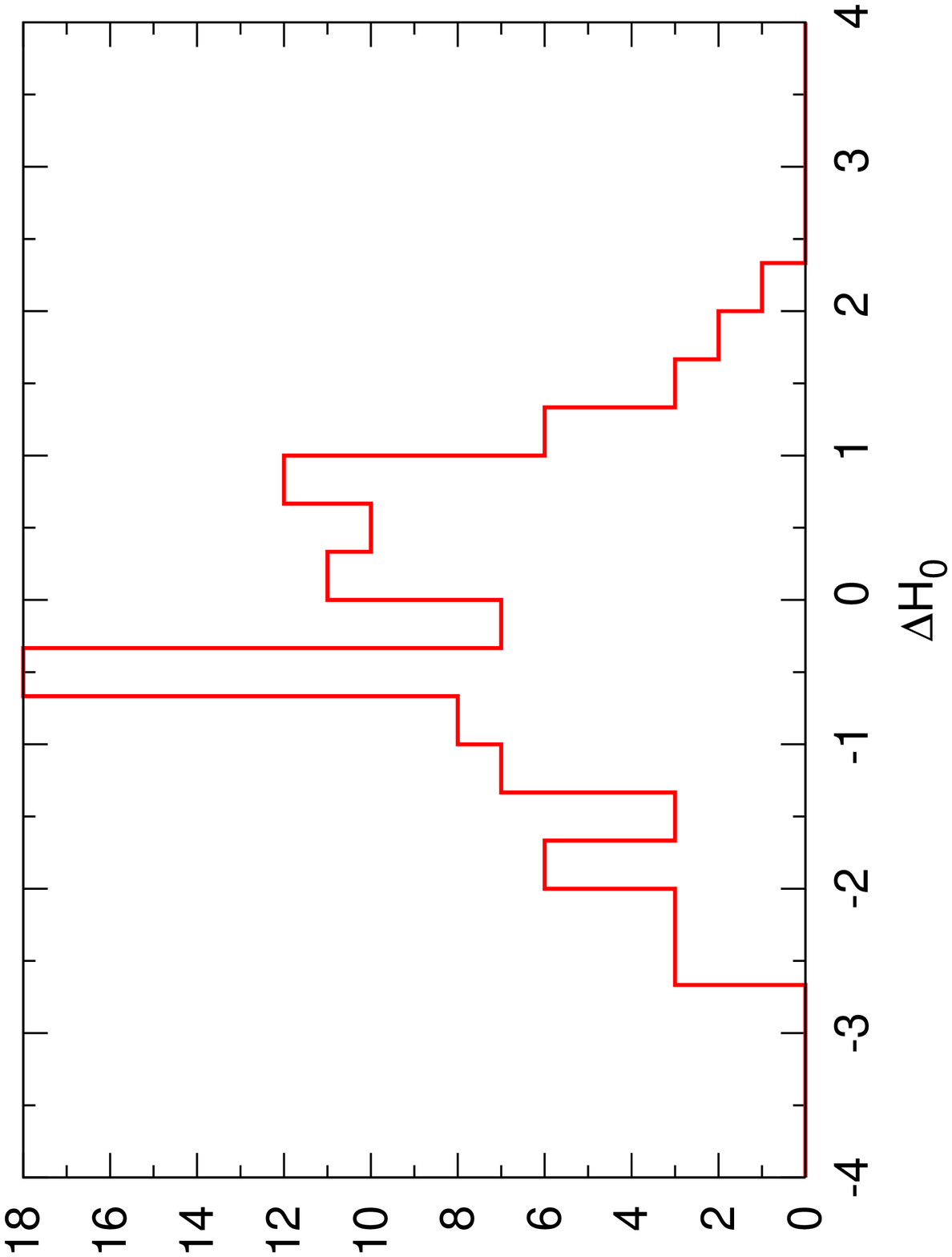}} \hspace{-7mm} 
  \subfloat[]{\incgraph{0.35}{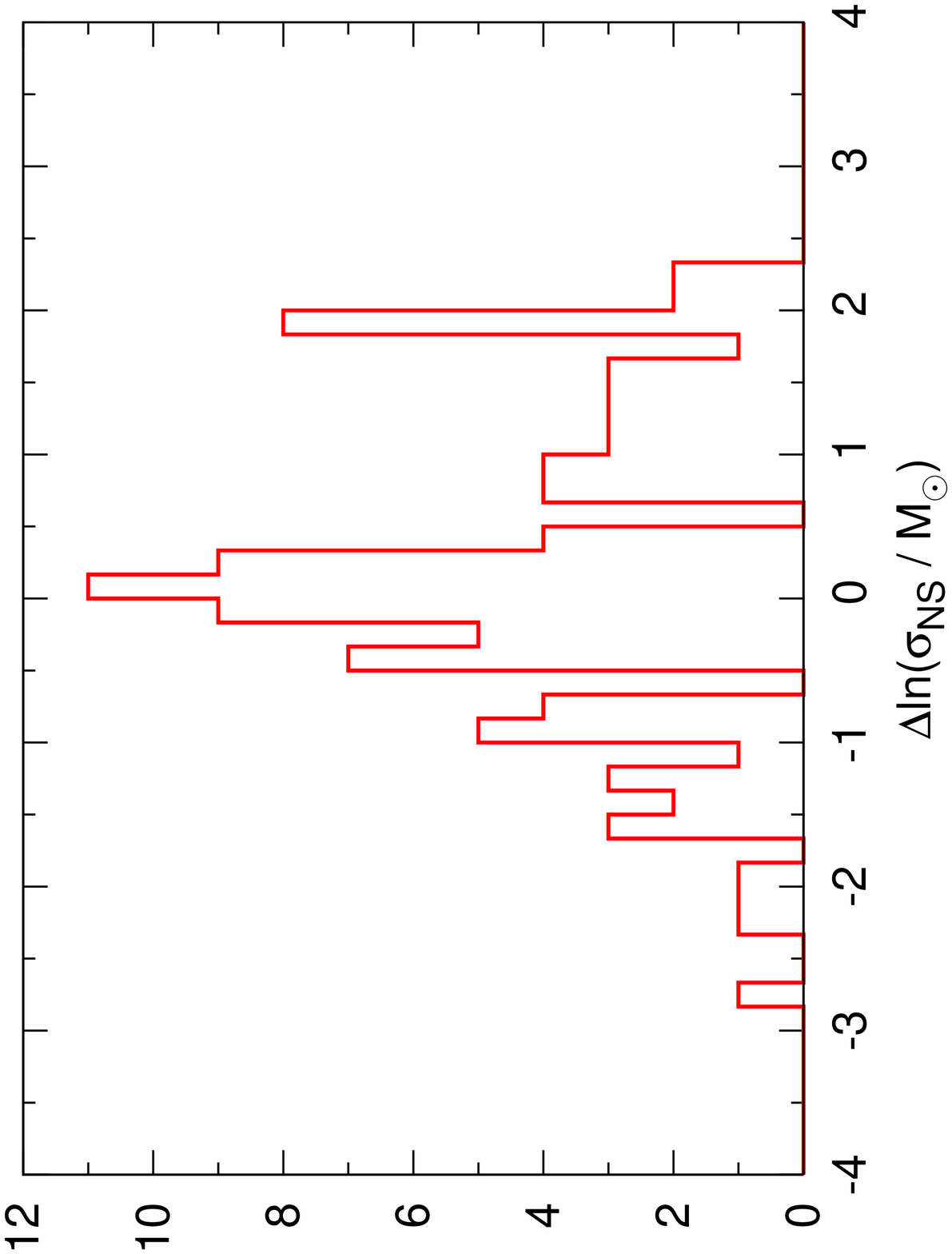}} \hspace{-7mm} 
  \subfloat[]{\incgraph{0.35}{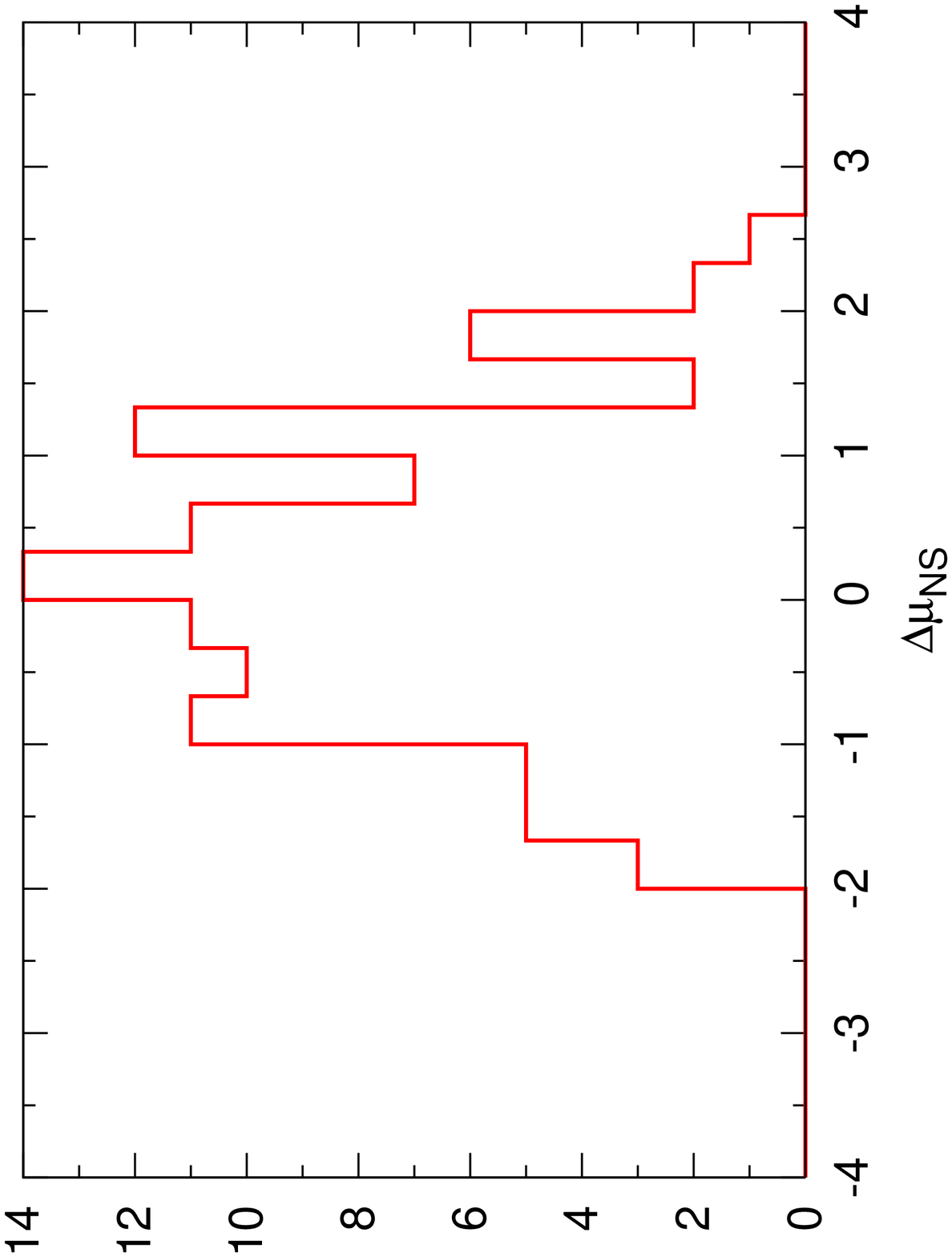}}
    \caption{\label{fig:rand-spread}Distribution of the Gaussian-fit standard deviations (top) and ``errors'' (bottom) of the recovered posteriors over 100 realizations, for $H_0$ (left), ${\ln}({\sigma}_{\rm{NS}})$ (center) and ${\mu}_{\rm{NS}}$ (right). More details are given in the text.}
\end{figure*}
To explore the spread in the best-fit parameters of the recovered posteriors over different realizations of the data catalog, we generated $100$ different realizations, keeping the intrinsic parameter values the same for each.

In each case, we fit a Gaussian to the $1$D posteriors and record the mean, ${\mu}$, standard deviation, ${\sigma}$, and the ``error'' in the mean. This last quantity is the number of standard deviations that the mean is offset from the intrinsic value, i.e.\ ${\Delta}X=({\mu}-X)/{\sigma}$, where $X$ is the value of the parameter used to generate the catalog \citep{gair2010}. A ${\pm}2{\sigma}$ offset encloses ${\sim}95{\%}$ of the Gaussian probability distribution, so we would reasonably expect most of the realizations to lie within this range.

Figure \ref{fig:rand-spread} shows the distributions of the Gaussian-fit standard deviations and ``errors'' for $H_0$, ${\ln}({\sigma}_{\rm{NS}})$ and ${\mu}_{\rm{NS}}$ over $100$ realizations of the AdLIGO-network data catalog. The distribution of the Gaussian-fit means for each parameter roughly resemble their respective posteriors, and the distribution of Gaussian standard deviations also appears approximately Gaussian. As we would have hoped, most of the realizations have a best-fit mean which is offset from the intrinsic value by less than $2{\sigma}$. As with the Gaussian-fit parameters, the ``error'' distribution is approximately Gaussian and centered around $0$ i.e.\ centered around the intrinsic value.

The most useful quantity here is the standard deviation of the reconstructed posterior distribution, as it characterizes how well we will be able to constrain the model parameters. The distribution over $100$ realizations displays the typical range of this ``measurement accuracy.'' Thus, ignoring measurement errors in the data, and with reference parameters used to generate the catalog, we could conceivably determine $H_0$, ${\sigma}_{\rm{NS}}$ and ${\mu}_{\rm{NS}}$ to an accuracy of ${\sim}{\pm}10$ km s$^{-1}$Mpc$^{-1}$, ${\sim}{\pm}0.004M_{\odot}$,\footnote{Evaluated using $\delta(\sigma_{\rm NS})=\sigma_{\rm NS}\times\delta(\ln{(\sigma_{\rm NS})})$, taking $\sigma_{\rm NS}$ to be the reference value and a typical error in $\ln{(\sigma_{\rm NS})}$ of $0.072$.} and ${\sim}{\pm}0.012M_{\odot}$ respectively.

\subsubsection{Including $\&$ accounting for errors}
\begin{figure*}
  \subfloat[]{\incgraph{0.35}{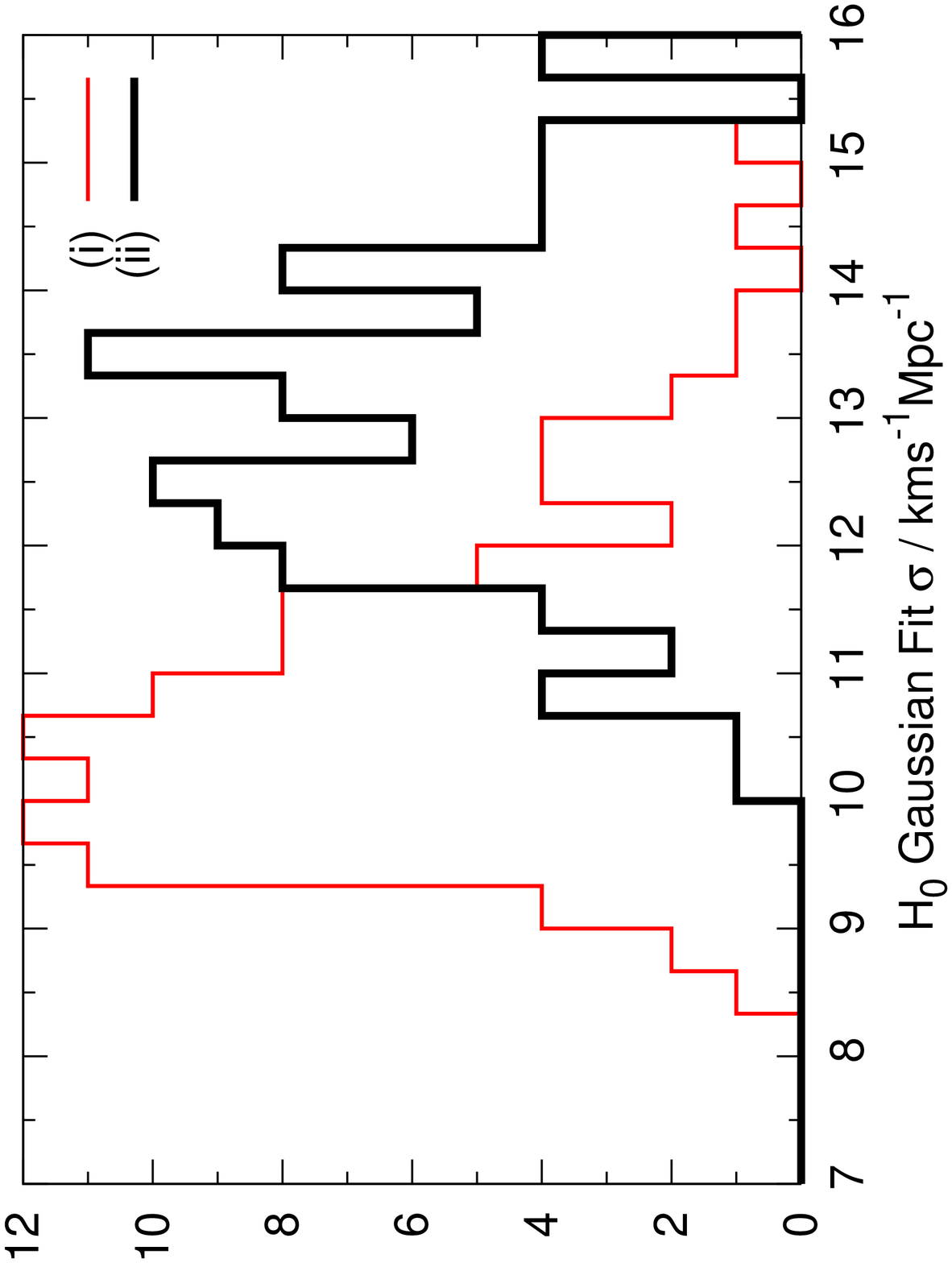}} \hspace{-7mm}                 
  \subfloat[]{\incgraph{0.35}{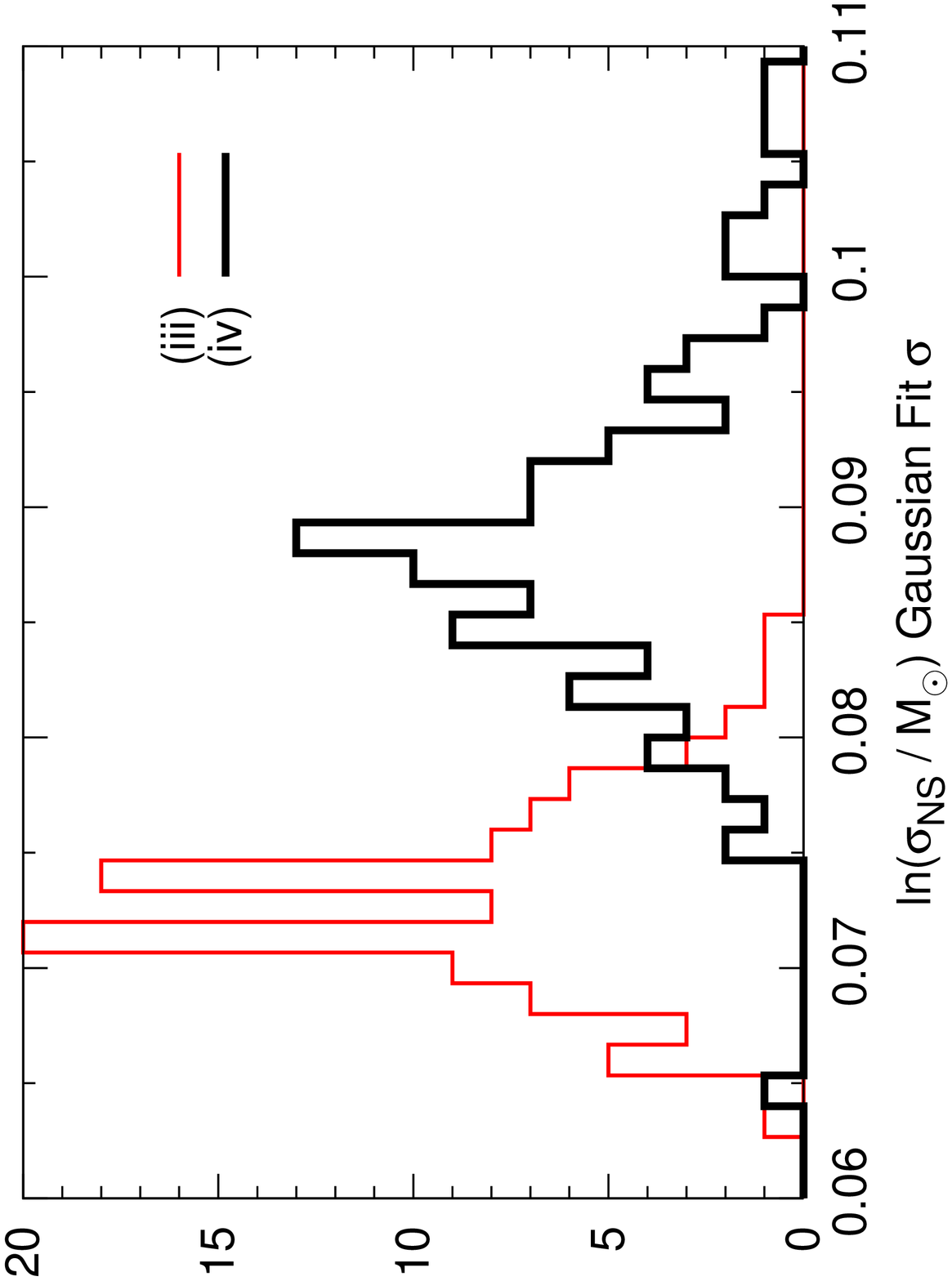}} \hspace{-7mm} 
  \subfloat[]{\incgraph{0.35}{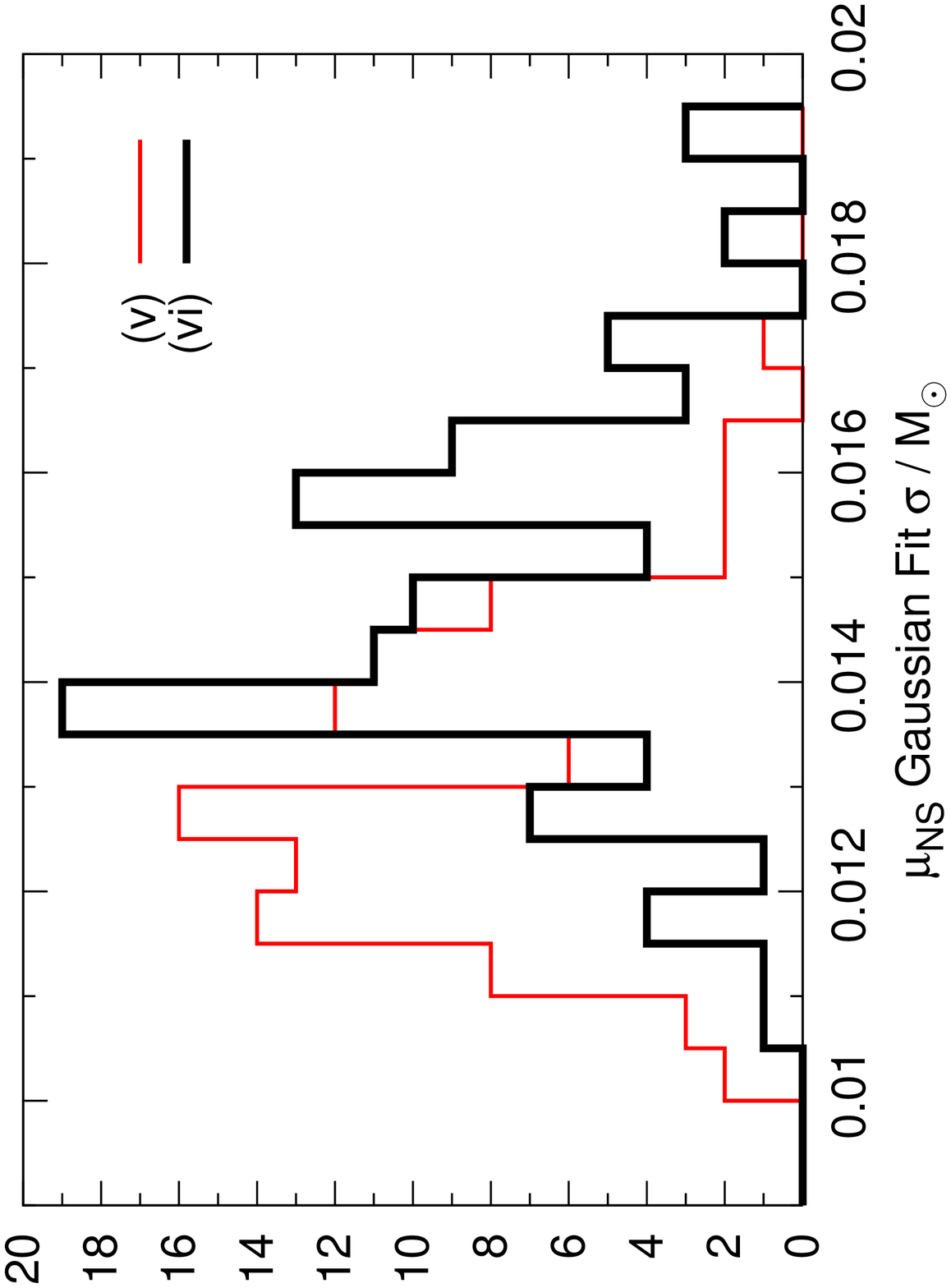}}
  \caption{\label{fig:error-compare}A comparison of the best-fit ${\sigma}$ distributions over $100$ realizations, between the case of no errors present in the data catalog, and the case of errors applied to system properties in the catalog. We have attempted to compensate for the errors in the data. (i), (iii) and (v) show the best-fit ${\sigma}$ distributions when no errors are applied to the system properties in the data catalog. (ii), (iv) and (vi) show the best-fit ${\sigma}$ distributions when the received data has errors.}
\end{figure*}

As discussed in Sec.\ \ref{sec:properties}, the system properties of each event in the catalog will include some error arising from instrumental noise. The data for each event will actually be in the form of posterior probability density functions (PDFs) for the properties, where previously we have assumed these are $\delta$-functions at the true values. We repeat the analysis assuming uncertainty in the source properties. We can include errors in the system properties in the data generation stage, by choosing the recorded values from a Gaussian distribution centered on the true value, with a standard deviation of $0.04{\%}$ for ${\mathcal{M}}_z$ and $({300/{\rho}}){\%}$ for $D_L$, where ${\rho}$ is the SNR of the detected event.

When we included errors in the data generation, but did not account for them in the analysis, we found that the model parameter posterior distributions were on average biased toward lower values of $H_0$, with biases also present in the ${\mu}_{\rm NS}$ and ${\sigma}_{\rm NS}$ distributions. When the errors are added, systems will move both to lower and to higher values of the luminosity distance. However, as we discussed in Sec.\ II, the sources at greatest distance have the most influence on our ability to measure the cosmology. We would therefore expect the sources shifted to greater distances to have most impact on the cosmological parameter estimation, biasing us toward smaller values of $H_0$, as we found. 


However, we can account for these errors in the analysis, by modifying the previous likelihood in Eq.\ (\ref{eq:cont-prob}) \citep{gair2010} to
\begin{align} \label{eq:error-integral}
{\mathcal{L}}(\overrightarrow{{\overrightarrow{\Lambda}}}|{\overrightarrow{\mu}},\mathcal{H})=&{e^{-N_{\mu}}}{\int}{\int}{\ldots}{\int}\left[p\left(\overrightarrow{n}={\overrightarrow{s}}-{\displaystyle\sum_{i}}{\overrightarrow{h_i}}({\overrightarrow{\lambda_i}})\right)\right. \nonumber\\
&\left.\times{\displaystyle\prod_{i=1}^{N_o}}r({\overrightarrow{{\lambda_i}}}|{\overrightarrow{\mu}})\right]d^{k}{\overrightarrow{\lambda_1}}d^{k}{\overrightarrow{\lambda_2}}{\ldots}d^{k}{\overrightarrow{\lambda_{N_o}}},
\end{align}
{\noindent}where, in our case, each system is associated with two cataloged properties such that $k=2$, and ${\overrightarrow{s}}$ is the detector output, which is a combination of $N_o$ signals, ${\overrightarrow{h_i}}$, and noise, $\overrightarrow{n}$. This is as an integral over all possible values of the source parameters that are consistent with the data. The first term inside the square bracket is the computed posterior PDF for the detected population of sources. Typical LIGO/Virgo DNS inspiral detections last only a few seconds, whilst AdLIGO/AdVirgo inspiral detections may be in-band for several minutes. Regardless, these detections should be uncorrelated, with independent parameter estimates \cite{mandel2010}, and so this first term reduces to the product of the posterior PDFs for each detection.

If the posterior PDF for a given source has been obtained via MCMC techniques, then the integral in Eq.~(\ref{eq:error-integral}) may be computed by summing over the chain samples. Thus, errors may be accounted for by making the following replacement in Eq.\ (\ref{eq:cont-prob}),
\begin{equation}
r({\overrightarrow{{\lambda}_i}}|{\overrightarrow{\mu}})\longrightarrow{\frac{1}{{\mathcal{N}}_i}}{\displaystyle\sum_{j=1}^{{\mathcal{N}}_i}}r({\overrightarrow{{\lambda}_i}}^{(j)}|{\overrightarrow{\mu}}),
\end{equation}
{\noindent}where ${\mathcal{N}}_i$ is the number of points in the chain for the $i^{\rm{th}}$ source's PDF, and ${\lambda}_i^{(j)}$ is the $j^{\rm{th}}$ element of the discrete chain representing this PDF. This technique does not assume a specific form for the PDF, and can be used in the case of multimodal distributions.\footnote{Multimodal distributions may result from partial degeneracies with other waveform parameters \citep{mandel2010}, such as the angular variables encapsulated in ${\Theta}$. Examples of this are shown in \cite{nissanke2010}, where the sky position of a detected system is pinned down, and the degeneracy between the inclination angle, ${\iota}$, and $D_L$ can lead to multimodal posteriors for $D_L$ which skew the peak to higher distances than the intrinsic value.}

In this analysis, we include errors on $D_L$ only, as those on the {\it redshifted} chirp mass ${\mathcal{M}}_z$ are very small and can be ignored.  (The uncertainty in the redshift estimate, which dominates the uncertainty in $H_0$ as discussed in Sec.\ \ref{sec:model}, arises from the width of the {\it intrinsic} chirp-mass distribution.)  We represent the $D_L$ posterior PDF for each source by a chain of $75$ points, drawn from a normal distribution with standard deviation ${\sigma}=(3/{\rho})D_L$, and a mean equal to the value in the data catalog, which in this analysis, as discussed earlier, includes an error to offset it from the true value. Whilst we adopt a simple Gaussian $D_L$ posterior PDF, the methodology we use here to account for errors is not reliant on the specific form of the PDF.

Using this analysis, we found that the bias in the posterior means for $H_0$, ${\sigma}_{\rm{NS}}$ and ${\mu}_{\rm{NS}}$ was corrected. In Fig.\ \ref{fig:error-compare} we show a comparison of the best-fit ${\sigma}$ distributions for each of the parameters when measurement errors are included (and accounted for), compared to the case in which they are ignored. It is clear that the presence of measurement errors decreases the measurement precision that we can achieve. However, the distributions overlap in all cases, and the peak of the error distributions is shifted only ${\sim}20{\%}$ higher. 

These errors only cause a shift in the measurement precision, so that we can ignore errors in the cataloged properties, with the knowledge that a full analysis would produce broadly the same results, but with ${\sim}20{\%}$ worse precision. The presence of errors (when accounted for) should therefore not affect our general conclusions about what a second-generation global network will be able to tell us about the underlying cosmological and source parameters.

\subsection{Dependence on number of observed events} \label{sec:merger-vary}
\begin{figure*}
  \subfloat[]{\incgraph{0.31}{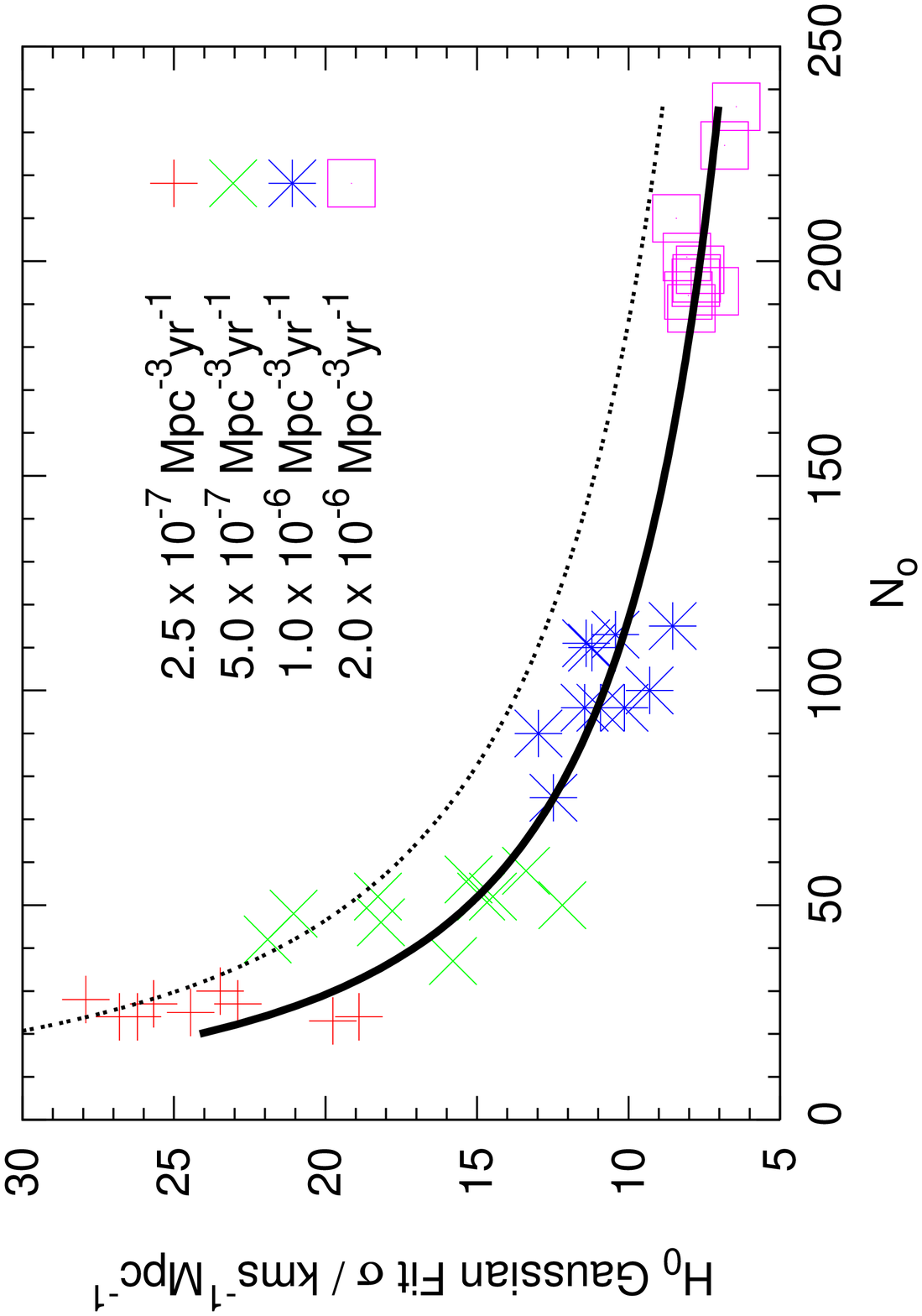}}                
  \subfloat[]{\incgraph{0.31}{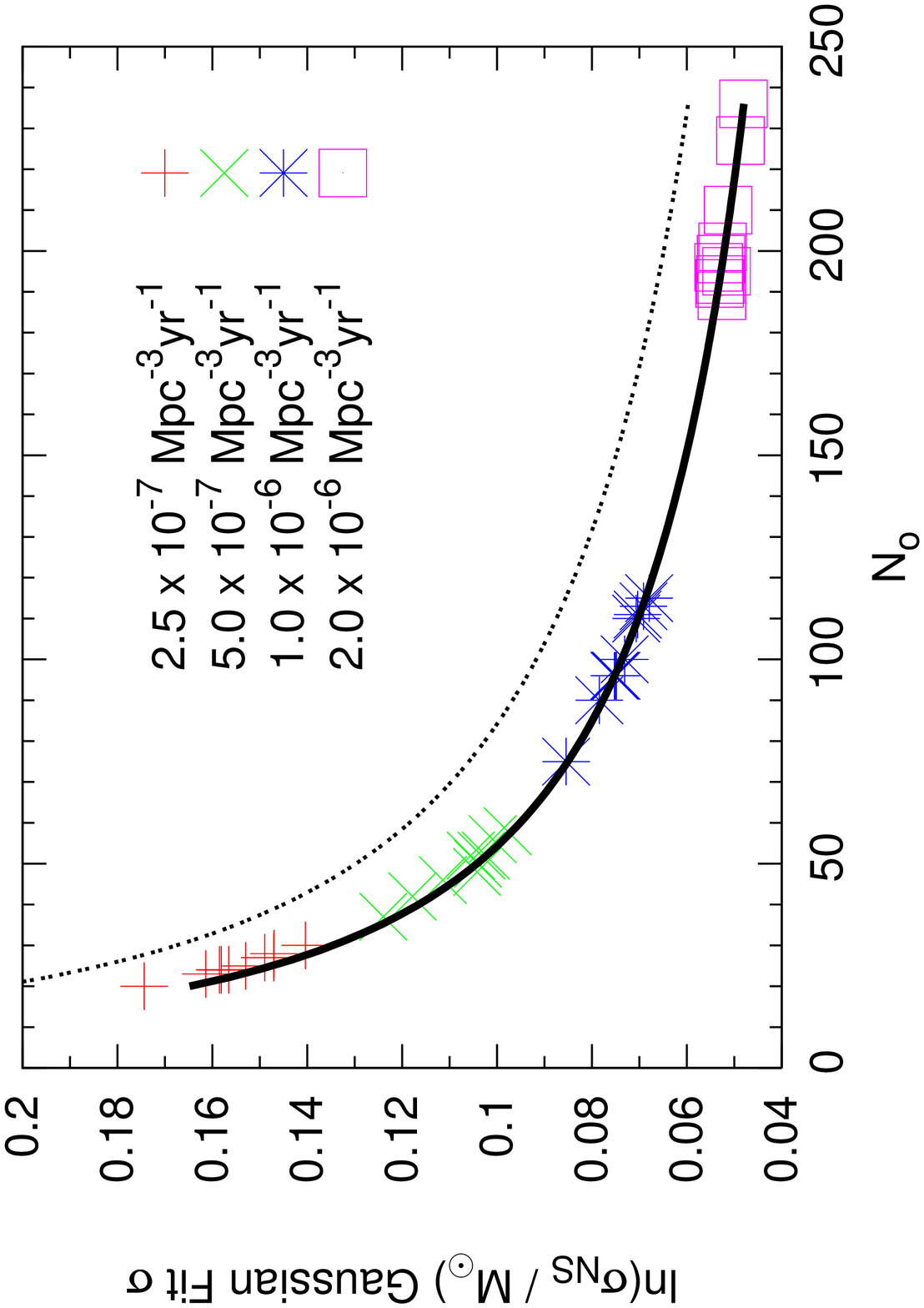}}
  \subfloat[]{\incgraph{0.31}{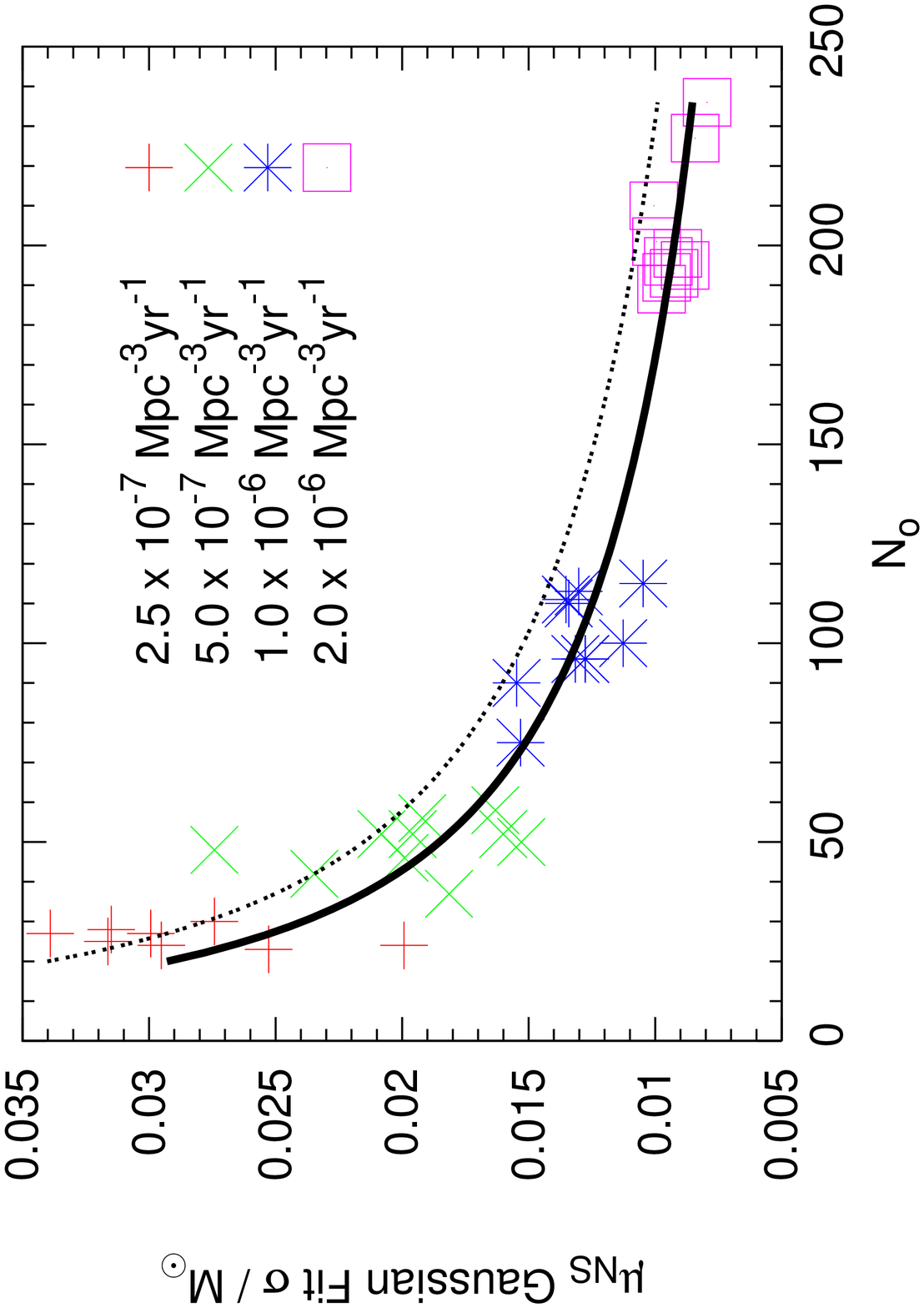}}
  \caption{\label{fig:num-vary}The ``measurement accuracy'' of each parameter (represented by the standard deviation of the Gaussian fit to the posterior) plotted against the number of observed events, $N_o$. The intrinsic parameters are kept fixed whilst the local merger-rate density, ${\dot{n}}_0$, is scaled up and down. The number of observed events scales linearly with the observation time and the local merger-rate density, such that the same result is achieved for twice the local merger-rate density if the observation time is halved. We see that a $1/{\sqrt{N_o}}$ relation is favored. The points and solid lines correspond to the case when we ignore errors, where the curves have gradients $108{\pm}2$ km s$^{-1}$Mpc$^{-1}$, $0.737{\pm}0.001$ and $0.131{\pm}0.002M_{\odot}$ respectively. The dashed lines are best-fit curves for the same analysis, but with measurement errors included and accounted for, for which the gradients are $136$ km s$^{-1}$Mpc$^{-1}$, $0.917$ and $0.152M_{\odot}$, respectively.}
\end{figure*}
\begin{table*}
\caption{The best-fit curves to the plots in Fig.\ \ref{fig:num-vary} are used to compute the percentage measurement precision of the model parameters. The local merger rates match the range quoted in \cite{abadie-rate2010}, where in our analysis, ${\dot{n}}_0=1.0$ Mpc$^{-3}$Myr$^{-1}$ gives ${\sim}100$ detections in $1$ year (at a network SNR threshold of $8$). In each case, the mean of the posterior distribution is taken at the reference value.\label{tab:accuracy-data}}
\begin{ruledtabular}
\begin{tabular}{c c c  c c  c c}
 \multirow{3}{*}{${\dot{n}}_0$ / Mpc$^{-3}$Myr$^{-1}$} & \multicolumn{6}{c}{Accuracy (${\sigma}_X/X$) / ${\%}$} \\ 
& \multicolumn{2}{c}{$H_0$} & \multicolumn{2}{c}{${\sigma}_{\rm{NS}}$} & \multicolumn{2}{c}{${\mu}_{\rm{NS}}$} \\ 
 & No errors & Errors & No errors & Errors & No errors & Errors \\
\hline


0.01 & 150 & 200 & 80 & 100 & 10 & 11 \\
1.0 & 15 & 20 & 7 & 9 & 1.0 & 1.1 \\
10.0 & 5 & 6 & 2 & 3 & 0.3 & 0.35 \\
50.0 & 2 & 3 & 1.0 & 1.3 & 0.14 & 0.16\\
\end{tabular}
\end{ruledtabular}
\end{table*}
The next question we will explore is how the measurement accuracy of the parameters depends on the number of cataloged events. This can be answered by changing the local merger-rate density, ${\dot{n}}_0$, or the observation time, $T$, whilst keeping the other model parameters fixed. We analyzed catalogs with different values of ${\dot{n}}_0$ around the previously used realistic value ($2.5{\times}10^{-7}$, $5.0{\times}10^{-7}$, $1.0{\times}10^{-6}$ and $2.0{\times}10^{-6}$ Mpc$^{-3}$yr$^{-1}$), using $10$ realizations in each case.

In Fig.\ \ref{fig:num-vary} we show the standard deviation of the recorded posterior distribution versus the number of cataloged events for each realization of each ${\dot{n}}_0$. The distributions are well fit by a function of the form,
\begin{equation}
{\sigma}_{X}\propto\frac{1}{\sqrt{N_o}},
\end{equation}
{\noindent}which one might expect; we have a population of $N_o$ events which we are using to statistically constrain a parameter, so we expect that the root-mean-squared error on the parameter should scale as ${\sim}1/{\sqrt{N_o}}$. The points and solid lines are the data and best-fit curves when we ignore measurement errors in the data generation, whilst the dashed lines are best-fit curves to data where we account for measurement errors, as in the previous section. 

Table \ref{tab:accuracy-data} shows the percentage fractional accuracy to which we could measure each model parameter, in both the case that we ignore errors and when we account for them. The range of local merger-rate densities reflects the quoted values in \cite{abadie-rate2010}, and the means of the posterior distributions are taken as the reference values.

The number of detected events will also depend on the SNR threshold, ${\rho}_{0,\rm{net}}$ \citep{oshaughnessy2010}. In practice, the network thresholds required for detection are often higher than ${\sim}8$ because a network performs more trials of the same data and is sensitive to both gravitational wave polarizations simultaneously. The result of increasing the threshold SNR to $10$ is to approximately halve the detection rate. If the expected detection rate is ${\sim}100$ yr$^{-1}$ in the ${\rho}_{0,\rm{net}}=8$ case, this becomes ${\sim}50$ yr$^{-1}$ in the ${\rho}_{0,\rm{net}}=10$ case.
\begin{figure} 
 \incgraph{0.45}{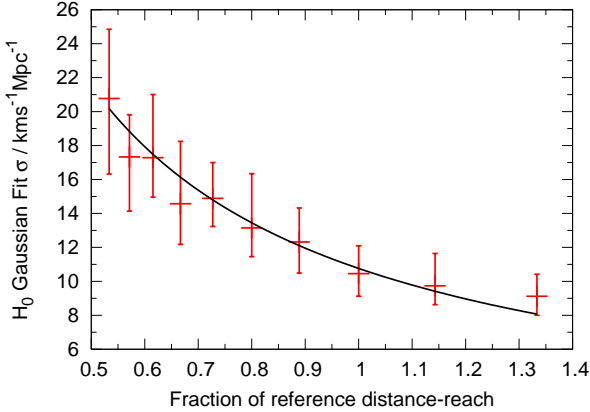}
    \caption{\label{fig:scale-reach}The variation of measurement accuracy with instrumental distance reach is shown. Each point represents the weighted mean of the $H_0$ measurement accuracy from $10$ realizations at a particular $r_{0,\rm{net}}$, where the error bars show the maximum and minimum values of $\sigma$ out of the $10$ values. All other parameters are at their reference values, and the total number of detections is scaled to $100$. The reference distance reach is $r_{0,\rm{net}}\sim176$ Mpc. The curve is a $(1/r_{0,\rm{net}})$ fit to the data, with gradient $10.8\pm0.2$ km s$^{-1}$Mpc$^{-1}$.} 
 \end{figure}

This halving of the detection rate is expected since,
\begin{equation}
\frac{V_{c,\rm{eff}}({\rho}_{0,\rm{net}}=10)}{V_{c,\rm{eff}}({\rho}_{0,\rm{net}}=8)}\simeq{\left(\frac{8}{10}\right)}^{3}=0.512,
\end{equation}
{\noindent}where $V_{c,\rm{eff}}$ is the effective comoving volume to which the network is sensitive.
We can achieve the same number of detections at higher SNR thresholds by increasing the observation time. Using a higher network SNR threshold is equivalent to assuming a lower characteristic distance reach for the network. By increasing ${\rho}_0$ to $10$, we cut the detection rate in half, which is roughly the decrease in the number of coincident detections when we shift from the HHL to HHLV network \cite{beau2008}. A network SNR threshold of $12$ reduces the detection rate to ${\sim}30$ yr$^{-1}$. 

To investigate the dependence of the $H_0$ measurement accuracy on the characteristic distance reach of the network (a prediction of our scaling arguments), we computed $10$ realizations at each of $10$ different network SNR thresholds, ranging from $6$ to $15$. The detection rates were kept the same at each SNR threshold by rescaling $\dot{n}_0$. The reference values were $\rho_{0,\rm{net}}=8$ and $r_{0,\rm{net}}=176$ Mpc, as used previously. At each $\rho_{0,\rm{net}}$, a weighted mean of the Gaussian-fit half-widths of the parameter posteriors was calculated, with error bars determined by the maximum and minimum half-widths out of the $10$ realizations. The results for $H_0$ are shown in Figure \ref{fig:scale-reach}. The fit favors a $(1/r_{0,\rm{net}})$ relationship, as expected from scaling arguments. There appears to be no effect on the measurement accuracy of the NS mass distribution parameters. No measurement errors were included on either the recorded $D_L$ or $\mathcal{M}_z$ values, and the detection rate was fixed in this analysis, so it is unsurprising that the measurement precision of the NS mass distribution parameters is unaffected by the reach of the network. In this particular investigation, given that the total number of events is unchanged, and therefore the number of masses to which we fit the NS mass distribution is unchanged, we do not expect the precision of the fit to change either.

This indicates that the measurement accuracies of ${\sigma}_{\rm{NS}}$ and ${\mu}_{\rm{NS}}$ quoted in this paper will be achievable at different $\rho_{0,\rm{net}}$ and $r_{0,\rm{net}}$ by scaling the observation times, or if the Universe has a different ${\dot{n}}_0$ than expected. However, the measurement accuracy of $H_0$ is also linked to the characteristic distance reach of the network.

\subsection{Dependence of measurement accuracy on intrinsic parameters} \label{sec:in-param}
\begin{figure*}
 \subfloat[]{\incgraph{0.31}{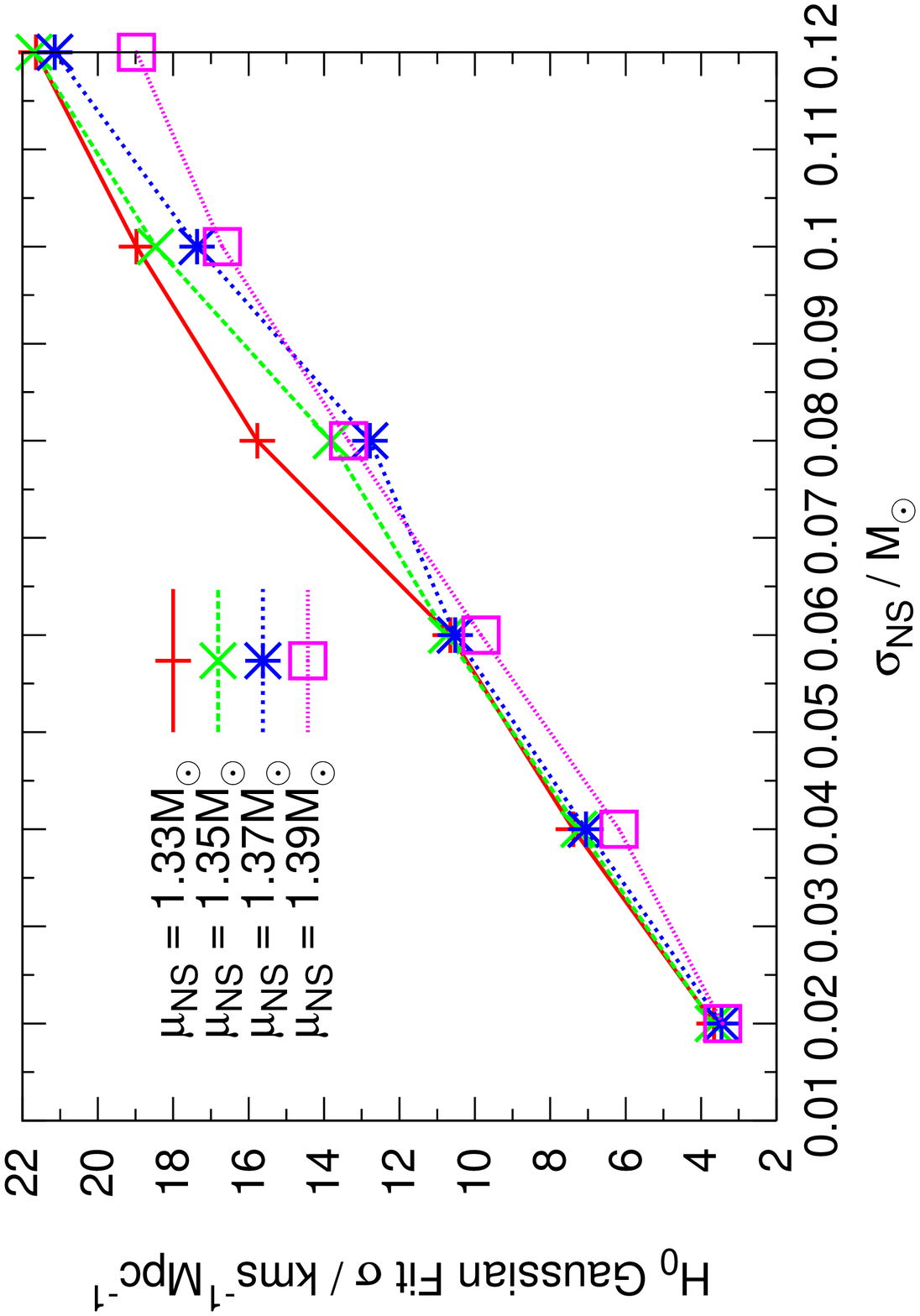}}              
  \subfloat[\label{fig:width-mean-precision}]{\incgraph{0.31}{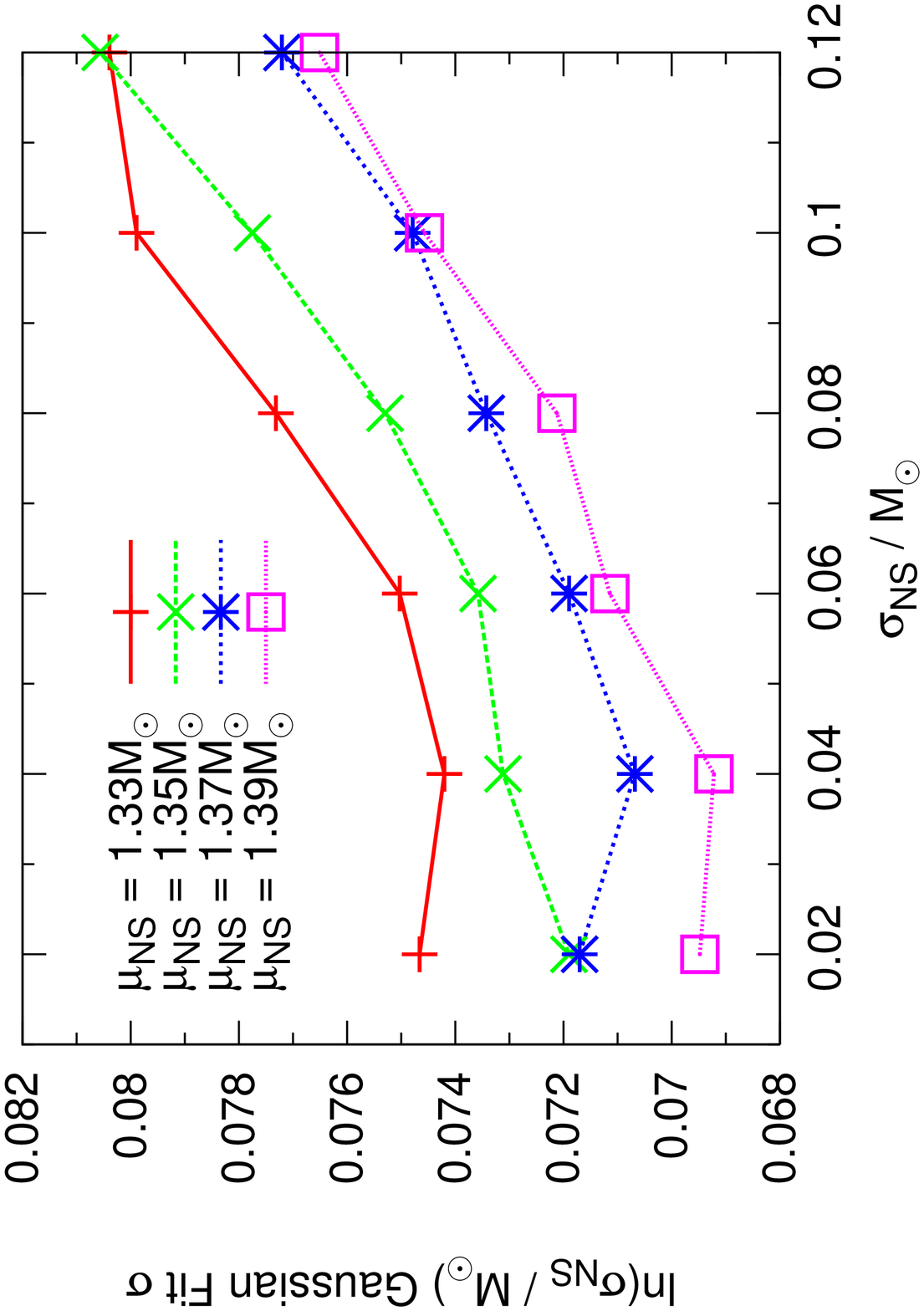}}
  \subfloat[]{\incgraph{0.31}{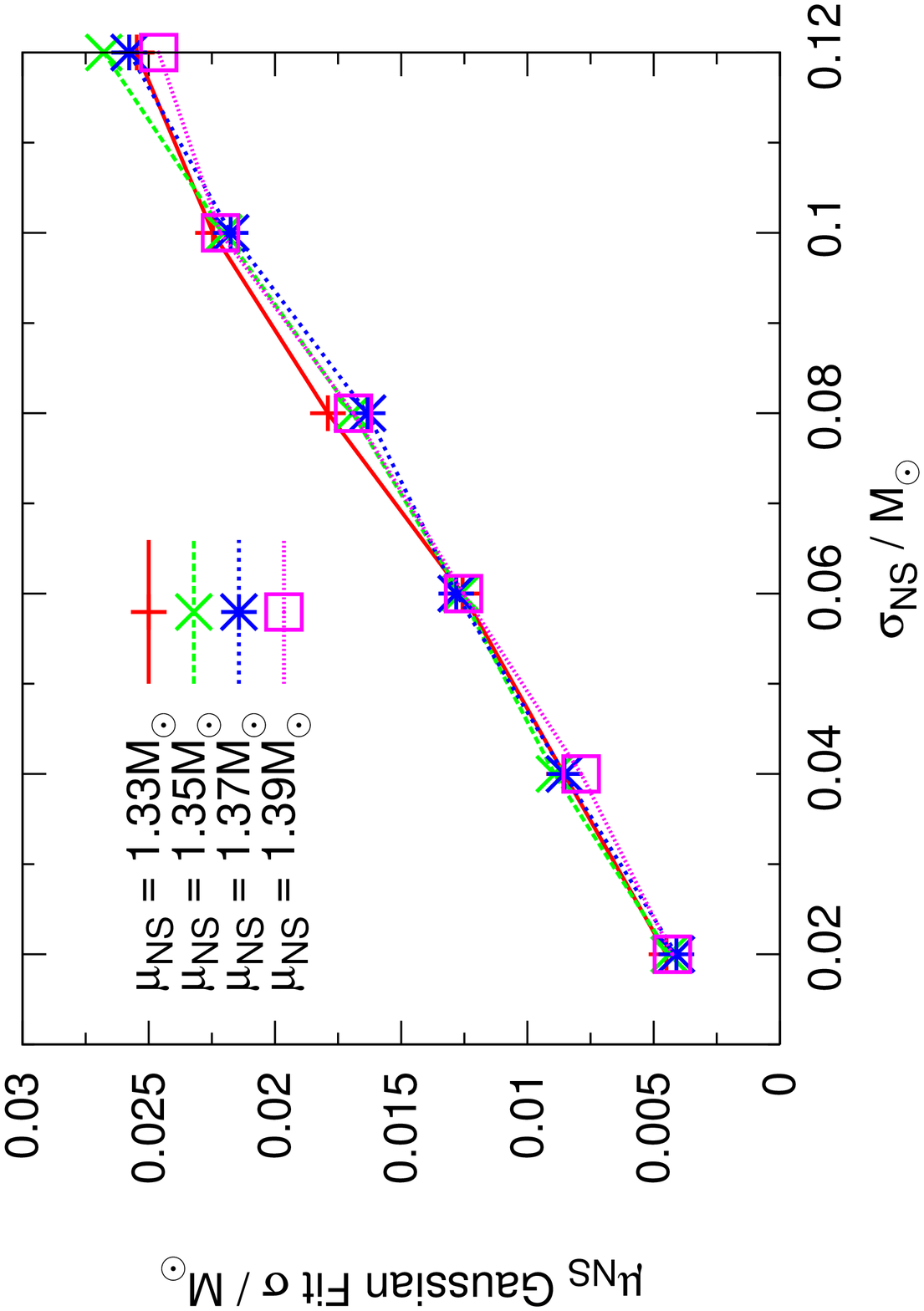}}
  \caption{\label{fig:param-vary-mean}The variation of the weighted mean (over $10$ realizations) of the Gaussian-fit standard deviations with the parameters of the underlying NS mass distribution. All other parameters are fixed at their reference values. The variation of the expected number of detections with ${\sigma}_{\rm{NS}}$ is less than one event, whilst for ${\mu}_{\rm{NS}}$ it is more significant. Thus all the posterior fit ${\sigma}$ values are scaled to the average number of detections for a given ${\mu}_{\rm{NS}}$ e.g.\ for ${\mu}_{\rm{NS}}=1.33M_{\odot}$ this average event number is ${\sim}95$, whilst for $1.39M_{\odot}$ it is ${\sim}110$, in $1$ year.}
\end{figure*}
It is also interesting to investigate how the constraints on the parameters of the underlying distributions depend on the parameters used to construct the distribution. This was done by generating $10$ data realizations at each of $24$ different combinations of the intrinsic ${\mu}_{\rm{NS}}$ and ${\sigma}_{\rm{\rm{NS}}}$. The intrinsic values of $H_0$, ${\Omega}_{m,0}$ and ${\alpha}$ were fixed at their reference values. The recorded measurement precision for a given intrinsic parameter combination was the weighted mean of this value over $10$ realizations.
Figure \ref{fig:param-vary-mean} shows the results of this analysis. One can see that the measurement precision depends on the width of the intrinsic NS mass distribution. An increase in the intrinsic ${\sigma}_{\rm{NS}}$ by a factor of $6$ leads to a reduction in the measurement accuracy on $H_0$ and ${\mu}_{\rm{NS}}$ by a factor of ${\sim}6$, but only leads to a modest $10{\%}$ reduction of the measurement accuracy for $\ln{({\sigma}_{\rm{NS}})}$.

The improvement of the measurement accuracy with a narrower intrinsic DNS mass distribution is a key result. In order to constrain the Hubble constant to within ${\sim}{\pm}10{\%}$ with ${\sim}100$ observations, we require the Gaussian half-width of the DNS mass distribution to be smaller than $0.04M_{\odot}$. The explanation for this is that we estimate the system chirp mass, $\mathcal{M}$, by dividing the redshifted chirp mass, ${\mathcal{M}}_z$, by $(1+z)$, where the $z$ is model-dependent (having been calculated from $D_L$ with given cosmological parameters). Thus, a narrower NS mass distribution will more effectively penalize model parameters which deviate from the intrinsic values. For ${\sigma}_{\rm{NS}}=0.13M_{\odot}$ \citep{kiziltan2010}, an accuracy of ${\sim}{\pm}10{\%}$ on $H_0$ would require ${\sim}O(1000)$ detections.

The dependence of the measurement precision on ${\mu}_{\rm{NS}}$ is not very clear from the left and right panels, but the effect on ${\sigma}_{\rm{NS}}$ is evident in Fig.\ \ref{fig:param-vary-mean}\subref{fig:width-mean-precision}. Varying the intrinsic ${\mu}_{\rm{NS}}$ from $1.33M_{\odot}$ to $1.39M_{\odot}$ provides a ${\sim}5-10{\%}$ gain in $\ln{({\sigma}_{\rm{NS}})}$ precision. The variation of the expected number of detections with ${\sigma}_{\rm{NS}}$ is less than one event, whilst for ${\mu}_{\rm{NS}}$ it is more significant. So, all the posterior fit ${\sigma}$ values were scaled to the average number of detections for a given ${\mu}_{\rm{NS}}$. This varies by ${\sim}15$ detections over the range of ${\mu}_{\rm{NS}}$ investigated. 

To explain the improvement in measurement precision with larger values of ${\mu}_{\rm{NS}}$, we note  Eq.\ (\ref{eq:snr}). We see that a larger mass distribution mean will, on average, imply larger individual NS masses. For a fixed SNR threshold, this allows detections to be made from larger $D_L$ values, thereby raising the effective comoving volume to which the network is sensitive. This raises the number of detections, and hence the parameter measurement accuracy. Rescaling all the measurement accuracies to $100$ events confirms that this is the dominant effect, as the different ${\mu}_{\rm{NS}}$ curves in Fig.\ \ref{fig:param-vary-mean} then overlap. Factorizing out the dependence on $N_o$ also confirms that the variation of the measurement accuracy with the width of the underlying NS mass distribution is a real feature.

Repeating the above analysis for fixed ${\mu}_{\rm{NS}}$, but with different combinations of $H_0$ and ${\sigma}_{\rm{NS}}$ confirms the variation of precision with ${\sigma}_{\rm{NS}}$. However, there appears to be no strong dependence on $H_0$ as it is varied by ${\pm}10$ km s$^{-1}$Mpc$^{-1}$ around the reference value.\footnote{The reference value is well constrained by WMAP+BAO+SNe analysis \citep{wmap_plus}.}
\subsection{Complementing GW data with GRB redshift data}
\begin{figure*}
  \subfloat[]{\incgraph{0.45}{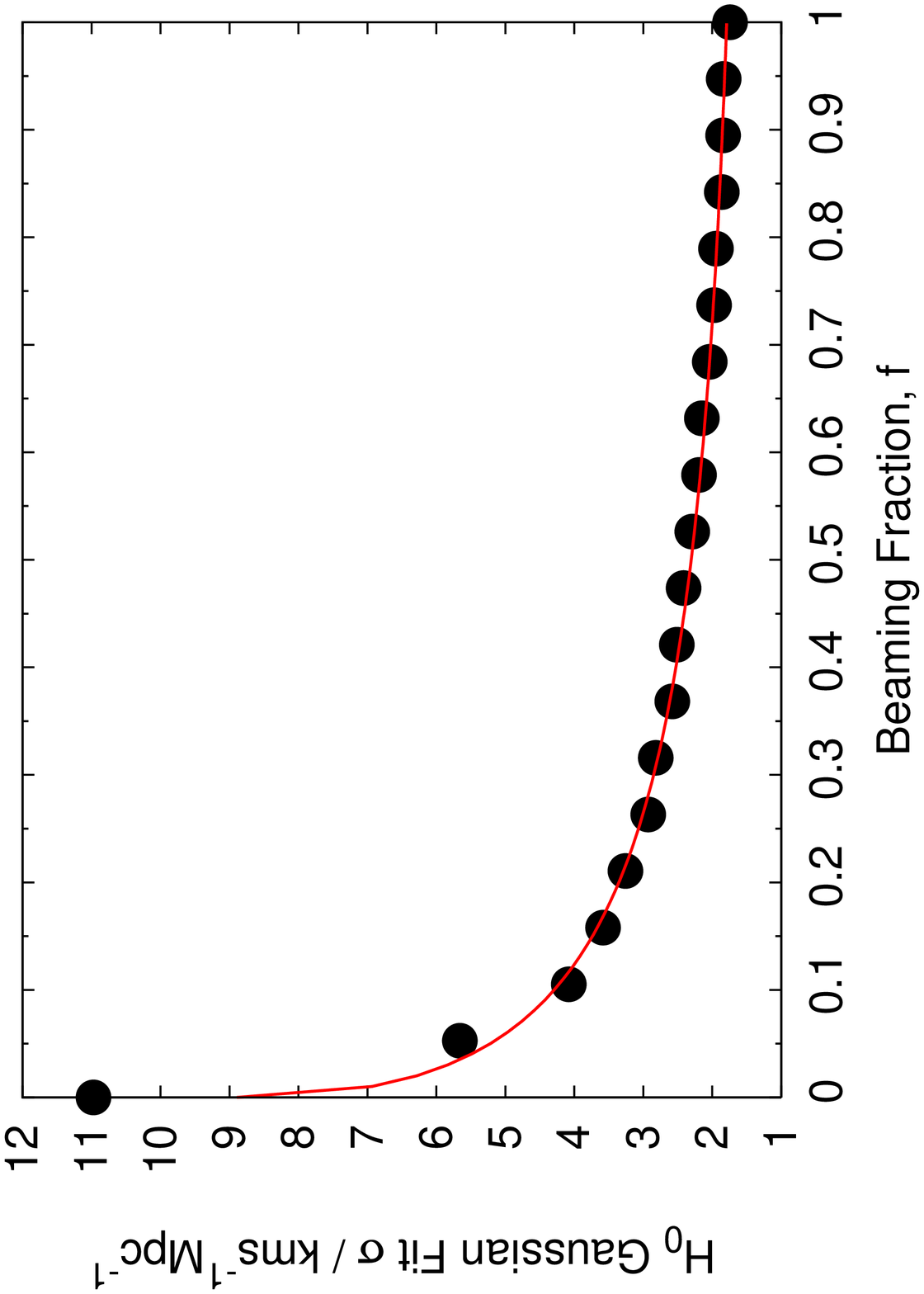}}                
  \subfloat[]{\incgraph{0.45}{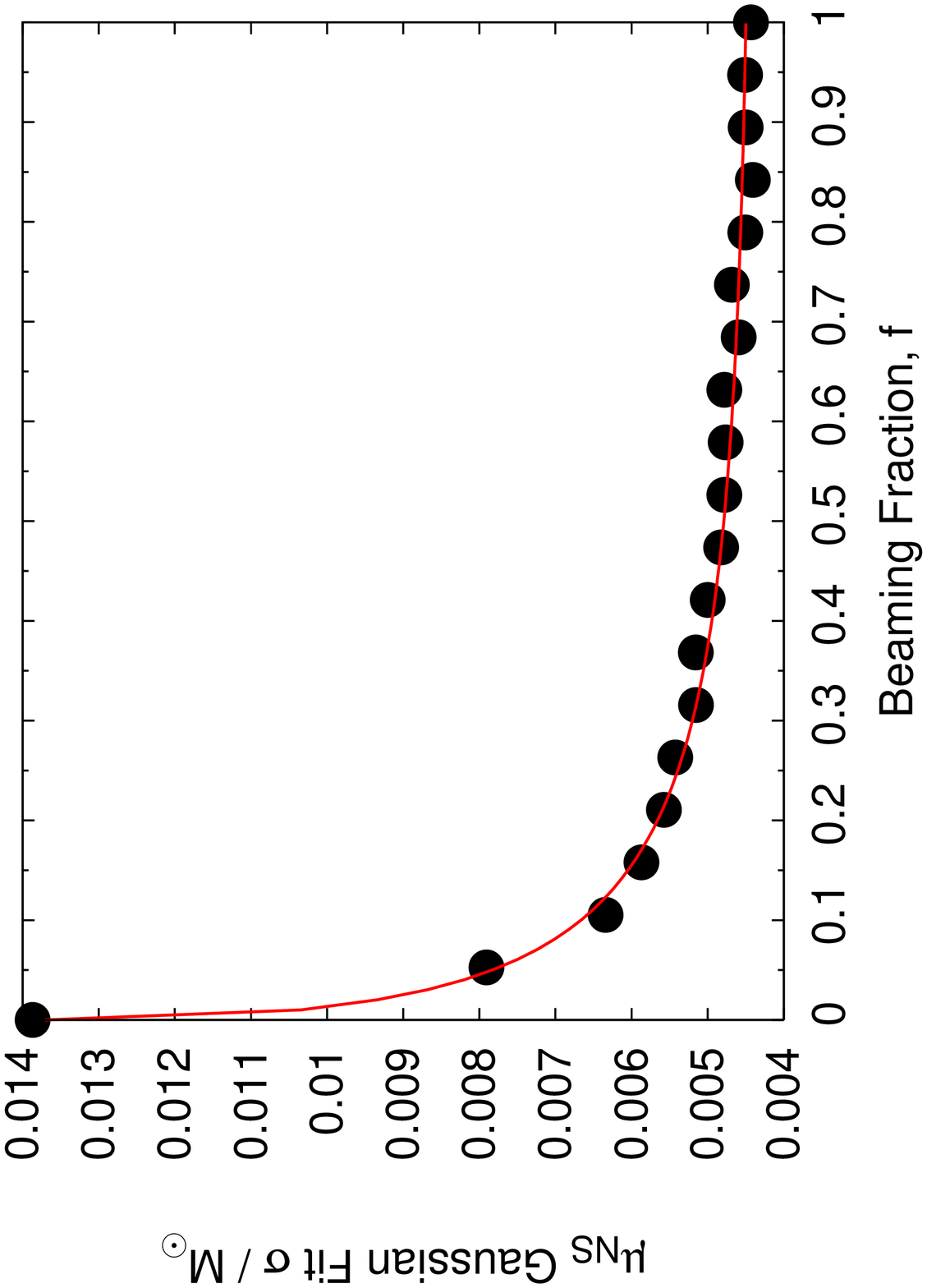}}
  \caption{ \label{fig:beam-frac}Plots of the measurement precision versus the beaming fraction of SGRBs, $f$. Approximately $fN_o$ events in the catalog were denoted as SGRBs, while the remaining ${\sim}(1-f)N_o$ events in the catalog were assumed to be GW-only events. A single data catalog was used repeatedly, with larger and larger fractions of it assumed to be observable as SGRBs. The data was generated with the reference parameters. The fitted curves are of the form $a\exp(-b\sqrt{x})+c$, where $(a,b,c)=(9.04,3.64,1.59)$ and $(0.00932,4.46,0.00439)$ for the $H_0$ and ${\mu}_{\rm{NS}}$ precisions respectively. The corresponding plot for ${\sigma}_{\rm{NS}}$ shows no obvious trend.}
\end{figure*}

In the above, we have assumed only the GW observations are available. However, if the redshift of the system is somehow known, then the background cosmological parameters can be directly probed using the luminosity distance, $D_L$, measured from the GWs \citep{Schutz86}.

Merging compact-object binary systems, such as NS-NS or NS-BH, are leading candidates for the explanation of short-duration gamma ray bursts (SGRB). SGRB events are among the most luminous explosions in the universe, releasing in less than one second the energy emitted by our Galaxy over one year \citep{rezzolla2011}, and involving intense outflows of gamma rays. There is therefore a good chance that EM counterparts to GW-detected DNS mergers will be observed. There is strong evidence that the emission from GRBs is not isotropic \citep{stanek1999,burrows2006,frail2001}, which may be due to the formation of relativistic jets in these systems \citep{rezzolla2011}. The redshift can be determined from the longer-wavelength SGRB afterglow \citep{metzger1997}.

Therefore, we may only observe the event electromagnetically if we happen to lie within the cone of the radiative outflow, whilst we should be able to detect the gravitational wave signal from any DNS merger within the AdLIGO-network horizon. We denote the beaming fraction, $f$, for SGRBs with a double-jet of opening angle ${\theta}_j$ by \citep{frail2001,dietz2011},
\begin{equation}
f=\frac{{\Omega}_{*}}{4{\pi}}=1-{\cos}\left(\frac{{\theta}_j}{2}\right).
\end{equation}
If we assume the population of SGRBs is randomly oriented on the sky, that their progenitors are all NS-NS mergers, that we will detect all SGRBs that are beamed toward us, and that the required SNR for a GW detection is independent of the existence of a counterpart, then the beaming fraction, $f$, is also the fraction of DNS inspiraling systems for which we would be able to gather redshift data. In practice, GW searches that are triggered by electromagnetic observations of SGRBs would have a greater distance reach than blind analyses, which would tend to increase the fraction of counterparts. However, gamma-ray telescopes operating in the advanced detector era might not have $100\%$ sky coverage, which would tend to reduce the fraction of counterparts. In addition, even with an SGRB counterpart we might not be able to determine the redshift, as this requires observation of an afterglow. However, all of the GW sources for advanced detectors will be at low redshift, for which the chances of measuring the redshift are significantly higher. In the following, when we refer to the beaming fraction, we will mean the fraction of GW detections with electromagnetically determined redshifts, which will be similar to the intrinsic beaming fraction, but not exactly the same for the reasons just described.

We performed a simple analysis to see how the measurement accuracy would improve if some fraction, $f$, of redshift data was available. A single data set was generated with the redshift, luminosity distance and redshifted chirp mass of each event recorded (with reference intrinsic parameters). The measured ${\mathcal{M}}_z$ and $D_L$ were drawn from Gaussian distributions centered at the true value, as described previously. However, as before the small error on ${\mathcal{M}}_z$ was ignored. When only a gravitational-wave signal is available the system properties are analyzed as previously, with the measured values assumed to be the true values. 

If an event is included in the SGRB fraction (with an associated redshift), then the likelihood is the product of the GW likelihood with the redshift posterior PDF, which we take to be a delta function (since spectroscopic redshift determination will be much more precise than GW determinations of $D_L$). The $D_L$ posterior was taken to be a Gaussian, centered around the measured value, and with a standard deviation of $30{\%}$ of this distance. This percentage error is a worst case, corresponding to a detection near the threshold SNR, i.e.\ $(300/{\rho}){\%}$ for $\rho=10$. Using a constant percentage of the distance as the width of the $D_L$ posterior is pessimistic, since closer events will be measured with greater accuracy. Integrating over redshift in Eq.\ (\ref{eq:error-integral}) picks out the value of the integrand at the true system redshift, since the redshift posterior is a delta function. Thus, the product in Eq.\ (\ref{eq:new-stat}) splits into two components, 
\begin{align} \label{eq:total-data}
\left[{\displaystyle\prod_{i=1}^{N_o}}{\frac{d^3N}{dtd{\mathcal{M}_z}d{D_L}}}{\bigg\vert_{i}}\right]{\hspace{-2.3mm}}\rightarrow{\hspace{-2.3mm}}&\left[{\displaystyle\prod_{j=1}^{fN_o}}\left({\frac{d^3N}{dtd{\mathcal{M}_z}d{z}}}{\bigg\vert_{j}}\right.\right.\nonumber\\
&\times\mathcal{N}\left[(D_L^{(j)}-D_L(z_j, {\overrightarrow{\mu}})), 0.3D_L^{(j)}\right]\Bigg)\Bigg]\nonumber\\
&\times\left[{\displaystyle\prod_{k=fN_o}^{N_o}}{\frac{d^3N}{dtd{\mathcal{M}_z}d{D_L}}}{\bigg\vert_{k}}\right].
\end{align}
The identification of the SGRB as an EM counterpart will vastly improve sky localization of the source, helping to beat down the degeneracies in the GW observations between $D_L$ and the inclination angle, ${\iota}$. A fuller investigation could also consider a prior on the inclination angle, given that the source is an SGRB with a collimated outflow \citep{nissanke2010,stanek1999,burrows2006}, and that emission has been observed. This would further help to improve the measurement precision of the luminosity distance.

If $f=1$, we find the precision for $H_0$ is ${\sim}2.0$ km s$^{-1}$Mpc$^{-1}$, compared to ${\sim}11.0$ km s$^{-1}$Mpc$^{-1}$ when $f=0$. The results are shown in Fig.\ \ref{fig:beam-frac}, along with fits to the data of the form $a\exp(-b\sqrt{x})+c$. The important result here is that the accuracy with which we are able to constrain $H_0$ and ${\mu}_{\rm{NS}}$ improves markedly with the beaming fraction. This is to be expected, since by recording $z$ and ${\mathcal{M}}_z$ we know exactly what the intrinsic chirp mass, ${\mathcal{M}}$, of the system is. The high accuracy of the redshift measurements restricts the space of model parameters through the Gaussian factor in Eq.\ (\ref{eq:total-data}). The same plot for ${\sigma}_{\rm{NS}}$ shows no trend at all. This may be because the measurement accuracy of ${\sigma}_{\rm{NS}}$ is most strongly linked to the number of cataloged events, rather than whether we include extra system information. 

This analysis could be sensitive to the errors we include in the data catalog, since the normal distribution in the left square-bracket of Eq. (\ref{eq:total-data}) will favor model parameters, $\overrightarrow{\mu}$, such that $D_L^{(j)}=D_L(z_j, {\overrightarrow{\mu}})$. $D_L^{(j)}$ is the mean of the $D_L$ posterior PDF for the $j^{\rm{th}}$ event, which may be skewed away from the true value. However, the intrinsic values were always consistent with the mean and width of the recovered posteriors, so this does not seem to be a significant problem.

The SGRB jet opening angle is poorly constrained by observations. In \cite{nakar2007}, the authors quote the inverse beaming fraction to be in the range, $1\ll f^{-1}<100$, giving $f\gtrsim10^{-2}$ or a jet opening angle ${\theta}_j\gtrsim16^{\circ}$, which is consistent with theoretical constraints on the jet half-opening angle~\cite{rezzolla2011}. Such models permit the jet half-opening angle to be as large as $30^{\circ}$, for which the beaming fraction becomes ${\sim}0.13$ \citep{rezzolla2011}. This would allow $H_0$ and ${\mu}_{\rm{NS}}$ to be measured with a precision more than twice that of their GW-only values (see Figure \ref{fig:beam-frac}). 

In \cite{nissanke2010}, the authors performed an analysis on multiple DNS inspirals detections in the AdLIGO-Virgo network with associated EM signatures. They assumed the sky location of the sources were known, and that $D_L$ and $z$ were measured, so that they directly probed the distance-redshift relation. With $4$ SGRBs they predicted $H_0$ could be measured with a fractional error of ${\sim}13{\%}$, improving to ${\sim}5{\%}$ for $15$ events. With $f=1$, and scaling the measurement precision as $1/\sqrt{N_o}$, we find $4$ SGRBs gives ${\sim}12.5{\%}$ precision, whilst $15$ gives ${\sim}6.5{\%}$ precision. Figure \ref{fig:beam-frac} indicates an $H_0$ precision of ${\sim}5\%$ when $f=0.15$; thus the second square bracket on the right of Eq.\ (\ref{eq:total-data}) slightly improves the measurement accuracy of $H_0$ compared to the first square bracket alone. These results are dependent on the modelled $D_L$ errors, but are broadly consistent with \cite{nissanke2010}. In contrast, we expect we can constrain $H_0$ to within ${\sim}{\pm}15{\%}$ using ${\sim}100$ GW events, with no EM signatures recorded for any of the GW detections. $15$ SGRB events out of the ${\sim}100$ GW events requires a beaming fraction of ${\sim}0.15$, which is rather optimistic given the current constraints on the jet opening angle. However this could conceivably be achieved over observation times longer than one year; additionally, the detection of an electromagnetic transient could allow the sensitivity volume to be increased in a triggered search.

In this section, we have not considered the possibility of redshift determination of the DNS inspiraling system via its association with a host galaxy. This could prove difficult in practice, since the sky error box is sufficiently large as to contain many candidate galaxies. In \cite{nuttall2010} the authors comment that over $100$ galaxies can be found in a typical LIGO/Virgo GW signal error box at a distance of $100$ Mpc. However, in the same work they introduced a ranking statistic which successfully imaged the true host of a simulated GW signal ${\sim}93{\%}$ of the time, if $5$ wide-field images were taken. The caveat here is that this statistic has only been tested out to $100$ Mpc, since comprehensive galaxy catalogs are lacking beyond this range. The catalog completeness is not $100{\%}$ at $100$ Mpc, and even if more distant, complete catalogs were available, the number of potential host galaxies in an AdLIGO/AdVirgo sky error box would be much greater. $D_L$ determination via network analysis may help to restrict the redshift range of these searches, but this is an area in need of future attention.

A novel method was proposed in \cite{macleod-hogan} in the context of LISA EMRI detections.
In that case, instead of precisely identifying the host galaxy of a GW detection (and thus the redshift of the source), the value of $H_0$ was averaged over all galaxies present in LISA's sky error box. 
Each galaxy in the box was weighted equally, and the chosen host galaxy was not included in the likelihood calculation to take into account the fact that the true host galaxy may not even be visible in available catalogs. They showed that sub-percent accuracies on $H_0$ would be possible if $20$ or more EMRI events are detected to $z\lesssim0.5$. This method has recently been investigated in the context of DNS inspirals in the advanced detector era, where a precision of a few percent on $H_0$ was claimed to be possible with $50$ detections \citep{delpozzo2011}. 

\section{Conclusions $\&$ Future Work} \label{sec:conclusion}

We have explored the capability of an advanced global network of GW interferometers, such as the AHL (Australia, Hanford, Livingston) or HHLV (Hanford, Hanford, Livingston, Virgo) configurations, to probe aspects of the background cosmology and the nature of the neutron-star mass distribution (for NSs in DNS systems). Current rate estimates suggest these systems could be a strong candidate for the first direct GW detection. With the reach of the advanced detectors, it may be possible to produce catalogs of tens of these systems along with their associated properties over the first few years of advanced detector operation.

We used a Bayesian theoretical framework to assess the posterior probability of the cosmological parameters and the mean and standard deviation of the NS mass distribution. Catalogs of DNS system mergers were generated, comprising the system redshifted chirp mass, ${\mathcal{M}}_z$ and luminosity distance, $D_L$, from which we endeavoured to statistically constrain the underlying parameters. 

We simulated catalogs of $100$ detected binaries (corresponding to a few years of observation for a local merger-rate density of $10^{-6}$ Mpc$^{-3}$yr$^{-1}$ \citep{abadie-rate2010}) for reference parameters $H_0=70.4$ km s$^{-1}$Mpc$^{-1}$, ${\Omega}_{m,0}=0.27$, ${\mu}_{\rm{NS}}=1.35M_{\odot}$, ${\sigma}_{\rm{NS}}=0.06M_{\odot}$, ${\alpha}=2.0$ (where ${\alpha}$ is the gradient of the redshift evolution of the NS-NS merger-rate density).  With such catalogs of detections we found it should be possible to measure the Hubble constant, as well as the mean and half-width of the DNS Gaussian mass distribution. $H_0$ should be constrained to ${\sim}{\pm}10$ km s$^{-1}$Mpc$^{-1}$, ${\ln}({\sigma}_{\rm{NS}})$ to ${\sim}{\pm}0.07$ and ${\mu}_{\rm{NS}}$ to ${\sim}{\pm}0.012M_{\odot}$. As a result of the restricted cosmological reach of second-generation detectors, ${\Omega}_{m,0}$ and $\alpha$ cannot be constrained by such observations. This is because the different cosmological density parameters do not significantly affect low $z$ luminosity distances, and low $z$ sources will not characterize the redshift evolution of the merger-rate density. 

The measurement accuracy was characterized by the width of a Gaussian fit to the recovered posterior distributions. We also attempted to account for measurement errors in the data catalog and found that taking errors into account would slightly broaden the recovered parameter distributions, but only by ${\sim}20{\%}$. This can be compensated for by longer observation times.

Keeping the intrinsic parameters fixed, and scaling the merger-rate density (or the observation time) allowed us to investigate how this precision varied with the number of cataloged events. We found that precisions varied as $N_o^{-1/2}$ for all three parameters. We also investigated the effect of changing the network SNR threshold, which has the same effect as reducing the distance reach of the network. Scaling the local merger-rate densities to give equal numbers of detections was enough to achieve the same precision on the NS mass distribution parameters, but the uncertainty in measuring $H_0$ also scales inversely with the distance reach of the network.

We also checked how the values of the intrinsic parameters themselves affected our ability to constrain them. Varying $H_0$ over a range of reasonable values had little impact on the measurement precision, but the effect of ${\sigma}_{\rm{NS}}$ was considerable. Changing ${\sigma}_{\rm{NS}}$ from $0.12M_{\odot}$ to $0.02M_{\odot}$ led to a factor of ${\sim}6$ increase in the precision on $H_0$ and ${\mu}_{\rm{NS}}$, but a modest ${\sim}10{\%}$ improvement on $\ln{({\sigma}_{\rm{NS}})}$. Our key result is that for $H_0$ to be constrained to within ${\sim}{\pm}10{\%}$ using ${\sim}100$ events (with the intrinsic $H_0$ and mean of the DNS mass distribution fixed at their reference values), then the half-width of the intrinsic DNS mass distribution would have to be less than $0.04M_{\odot}$.

Finally, considering that NS-NS and NS-BH merger events are leading candidates for the progenitors of short-duration gamma-ray bursts \citep{rezzolla2011,dietz2011}, we investigated how the measurement precision would improve if redshift data were available for some fraction of the catalog. The redshift could be deduced from the afterglow of the SGRB or from the closest projected galaxy. The fraction of GW detections that have observable EM counterparts will depend on the opening angle of the SGRB jets. The most recent GR-MHD simulations permit a half-opening angle of $30^{\circ}$, for which the maximum fraction of the DNS inspiraling systems that could have an observable EM is ${\sim}0.13$ \citep{rezzolla2011} (this fraction could be further increased by the greater sensitivity of GW searches triggered on EM transients). This would permit a significant improvement on the measurement precision of $H_0$ and ${\mu}_{\rm{NS}}$ to more than double their GW-only precisions. There appears to be no effect on the measurement precision of ${\sigma}_{\rm{NS}}$.

Our results were based on a single-interferometer formalism to describe the global network, assuming that LIGO-Australia would be nearly antipodal to the U.S sites and have identical sensitivity. There is no difference in the number of expected detections between the HHL and AHL configurations, although a slight improvement in the detection efficiency is expected for the HHLV network \citep{fairhurst2011}. We can penalize the number of coincident detections made by all interferometers by raising the network SNR threshold from $8$ to $10$. This cuts the detection rate in half, but this can be compensated for by longer observation times.

We have shown the significant potential for a network of second-generation detectors to provide an independent measurement of the Hubble constant, and to determine the neutron-star mass distribution for those NSs found in DNS systems. 
Even  more powerful constraints should be possible with the \textit{Einstein Telescope} (ET), a proposed third generation ground-based interferometer with an arm-length of $10$ km \citep{punturo2010b}.

ET will be sensitive to sources out to $z\sim2$ for DNS inspirals, with the expected number of detections in one year being ${\sim}O(10^{5}-10^{6})$ \citep{punturo2010a,punturo2010b}. When we compare this to AdLIGO's $445$ Mpc reach, giving ${\sim}100$ network detections, we see the clear improvement ET will offer. With such a large reach and detection rate we anticipate a much greater measurement precision on $H_0$, as well as the other parameters discussed in this paper. For the model parameters at the reference values, and using our analysis to extrapolate for a conservative ET detection rate gives ${\sim}0.5{\%}$, ${\sim}0.3{\%}$ and ${\sim}0.03{\%}$ precision on $H_0$, ${\sigma}_{\rm{NS}}$ and ${\mu}_{\rm{NS}}$ respectively. Furthermore, the cosmological reach of ET may permit ${\Omega}_{m,0}$ and ${\Omega}_{k,0}$ to be constrained, with consequences even for probing the dark energy equation of state parameter, $w$ \cite{et-cosmography,et-dark-energy}.  Preliminary results for our future ET analysis has constrained ${\Omega}_{m,0}$ to ${\sim}{\pm}30{\%}$ with $100$ events (where, for this preliminary study, we used the same methodology as in the present paper, but with the characteristic distance reach modified to account for the sensitivity curve of the early ET design study, ET-B \cite{et-b-sensitivity}), and scaling this for a conservative ET detection rate of $10^5$ yr$^{-1}$ gives ${\sim}{\pm}0.9{\%}$. The ability to detect $z\gtrsim 2$ events may provide an opportunity to measure the evolution of the DNS merger-rate density, which will shed light on the evolution of the star-formation rate. Some of the techniques used in this paper would have to be adapted for any ET analysis, e.g., the approximation to deduce the redshift from the luminosity distance would have to be replaced by the full root-finding algorithm. This coupled with the huge number of catalogued events would lead to longer computation times.

In this analysis we have considered a global second-generation GW-interferometer network. The improvement offered by a Southern Hemisphere gravitational-wave detector would be significant for sky localization (though only moderate for distance estimates), but this may not be realized. However, even with the HHLV network, we will still be able to place constraints on the underlying model parameters by overcoming the reduced coincident detection rate with a longer duration network science run. For now, we have shown that if a global network is successful in detecting populations of inspiraling DNS systems, then gravitational wave astronomy can begin to place independent and interesting constraints on $H_0$, as well as the neutron-star mass distribution. This will be a step toward using gravitational-wave astronomy for precision astrophysics.

\begin{acknowledgments}
S.R.T is supported by the STFC. J.R.G is supported by the Royal Society.  I.M was supported by the NSF and is also grateful for the hospitality of AEI Golm and AEI Hannover.  We thank Rai Weiss, Stas Babak and Drew Keppel for useful discussions.

\end{acknowledgments}

\bibliography{refs}

\end{document}
%